\documentclass[preprint,3p]{elsarticle/elsarticle}

\usepackage{stmaryrd}
\usepackage{amsmath}
\usepackage{amssymb}
\usepackage{amsthm}
\usepackage{centernot}
\usepackage{bussproofs}
\usepackage{tikz}
\usepackage{hyperref}
\usepackage{scrextend}
\usepackage{semantic}
\usepackage{changebar}
\usepackage{combelow}

\pdfoutput=1

\title{Flag-Based Big-Step Semantics}

\author{Casper~Bach~Poulsen\corref{cor}\fnref{fn1}}
\ead{cscbp@swansea.ac.uk}

\author{Peter~D.~Mosses\corref{cor}}
\ead{p.d.mosses@swansea.ac.uk}

\address{
  Department of Computer Science\\
  Swansea University\\
  Singleton Park, Swansea\\
  SA2 8PP, UK}

\cortext[cor]{Corresponding author}

\fntext[fn1]{Present address: Programming Languages Group, Delft
  University of Technology, Mekelweg 4, 2628 CD Delft, Netherlands}

\journal{Journal of Logical and Algebraic Methods in Programming}

\biboptions{longnamesfirst,square,semicolon,numbers}

\mathlig{||d}{\shortdownarrow}
\mathlig{||u}{\shortuparrow}
\mathlig{-inf->}{\overset{\infty}{\rightarrow}}
\mathlig{=inf=>}{\overset{\infty}{\Rightarrow}}
\mathlig{=del=>}{\overset{\mathsf{d}}{\Rightarrow}}
\mathlig{=co=>}{\overset{\mathsf{co}}{\Rightarrow}}
\mathlig{=co=v}{\Downarrow^{\mathsf{co}}}
\mathlig{=du=v}{\Downarrow^{\mathsf{du}}}
\mathlig{=du=>}{\overset{\mathsf{du}}{\Rightarrow}}
\mathlig{=Nco=>}{\underset{\mathsf{N}}{\overset{\mathsf{co}}{\Rightarrow}}}
\mathlig{=delco=>}{\overset{\mathsf{dco}}{\Rightarrow}}
\mathlig{=E=>}{\Rightarrow_{\textsc{e}}}
\mathlig{=B=>}{\Rightarrow_{\textsc{b}}}
\mathlig{=NU=>}{\Rightarrow_{\textsc{nu}}}
\mathlig{=*=>}{\overset{*}{\Rightarrow}_{\textsc{nu}}}
\mathlig{=G=>}{\Rightarrow_{\textsc{f}}}
\mathlig{=GE=>}{\Rightarrow_{\textsc{fe}}}
\mathlig{=coG=>}{\overset{\mathsf{co}}{\Rightarrow}_{\textsc{f}}}
\mathlig{=duG=>}{\overset{\mathsf{du}}{\Rightarrow}_{\textsc{f}}}
\mathlig{=I=>}{\Rightarrow_{\textsc{i}}}
\mathlig{=IE=>}{\Rightarrow_{\textsc{ie}}}
\mathlig{=duIG=>}{\overset{\mathsf{du}}{\Rightarrow}_{\textsc{if}}}
\mathlig{=IEG=>}{\Rightarrow_{\textsc{ief}}}
\mathlig{=IT=>}{\Rightarrow_{\textsc{it}}}
\mathlig{=coIT=>}{\overset{\mathsf{co}}{\Rightarrow}_{\textsc{it}}}
\mathlig{=IA=>}{\Rightarrow_{\textsc{ia}}}
\mathlig{=IEA=>}{\Rightarrow_{\textsc{iea}}}
\mathlig{=IB=>}{\Rightarrow_{\textsc{ib}}}
\mathlig{=IEB=>}{\Rightarrow_{\textsc{ieb}}}
\mathlig{=SB=>}{\Rightarrow_{\textsc{sb}}}
\mathlig{=SEB=>}{\Rightarrow_{\textsc{seb}}}
\mathlig{-SB->}{\to_{\textsc{sb}}}
\mathlig{~>}{\leadsto}
\mathlig{|->}{\mapsto}
\mathlig{-fin->}{\xrightarrow{\mathrm{fin}}}
\mathlig{->*}{\to^*}
\mathlig{==>}{\Longrightarrow}
\mathlig{[<}{\langle}
\mathlig{>]}{\rangle}
\mathlig{|>}{\triangleright}
\mathlig{=v}{\Downarrow}
\mathlig{-co->s*}{\overset{\mathsf{co}}{\to}^\star}
\mathlig{->s*}{\to^\star}
\mathlig{-du->s*}{\overset{\mathsf{du}}{\to}^\star}

\setpremisesend{1pt}
\setnamespace{1pt}

\newtheorem{definition}{Definition}

\newtheorem{lemma}[definition]{Lemma}
\newtheorem{theorem}[definition]{Theorem}

\newtheorem{example}[definition]{Example}

\reservestyle{\syn}{\textsf}
\syn{exc,seq2,abort,div,conv,skip,do-while,throw,while,if,ref,deref,dealloc,input,catch}

\reservestyle{\ann}{\textsf}
\syn{co,du}

\newcommand{\hibox}[1]{\fcolorbox{gray!40}{gray!40}{#1}}

\begin{document}

\begin{abstract}
  Structural operational semantic specifications come in different
  styles: small-step and big-step. A problem with the big-step style
  is that specifying divergence and abrupt termination gives rise to
  annoying duplication. We present a novel approach to representing
  divergence and abrupt termination in big-step semantics using status
  flags. This avoids the duplication problem, and uses fewer rules and
  premises for representing divergence than previous approaches in the
  literature.
\end{abstract}

\maketitle

\section{Introduction}
\label{sec:introduction}
Formal specifications concisely capture the meaning of programs and
programming languages and provide a valuable tool for reasoning about
them. A particularly attractive trait of \emph{structural
  specifications} is that one can prove properties of programs and
programming languages using well-known reasoning techniques, such as
induction for finite structures, and coinduction for possibly-infinite
ones.

In this article we consider the well-known variant of structural
specifications called \emph{structural operational semantics} (SOS)
\cite{Plotkin2004astructuralapproach}. SOS rules are generally
formulated in one of two styles: \emph{small-step}, relating
intermediate states in a transition system; and \emph{big-step} (also
known as \emph{natural semantics} \cite{Kahn87naturalsemantics}),
relating states directly to final outcomes, such as values, stores,
traces, etc. Each style has its merits and drawbacks. For example,
small-step is regarded as superior for specifying interleaving,
whereas big-step is regarded as superior for compiler correctness
proofs
\cite{Leroy2009coinductivebigstep,Nipkow2014concretesemantics,Hutton2007whatis}
and for efficient interpreters
\cite{Danvy2004refocusingin,BachPoulsen2013generatingspecialized}. Different
styles can also be used for specifying different fragments of the same
language.

Big-step SOS rules, however, suffer from a serious \emph{duplication
  problem} \cite{Chargueraud2013prettybigstep}. Consider, for example,
the following rule for sequential composition:
\[
  \inference{
    ( c_1,\sigma ) => \sigma'
    &
    ( c_2,\sigma' ) => \sigma''
  }{
    ( c_1; c_2, \sigma ) =>
    \sigma''
  }[B-Seq]
\]
This rule is \emph{inductively} defined and covers the sequential
composition of all \emph{finite} computations for $c_1$ and $c_2$. But
what if either $c_1$ or $c_2$ is an \emph{infinite} computation, i.e.,
\emph{diverges}? The traditional approach to representing this in
big-step SOS is to introduce a separate, \emph{coinductively} defined,
relation $=inf=>$:
\[
  \inference{
    ( c_1,\sigma ) =inf=>
  }{
    ( c_1; c_2, \sigma ) =inf=>
  }[B-$\infty$-Seq1]
  \qquad
  \inference{
    ( c_1,\sigma ) => \sigma'
    &
    ( c_2,\sigma' ) =inf=>
  }{
    ( c_1; c_2, \sigma ) =inf=>
  }[B-$\infty$-Seq2]
\]
Here, the first premise in B-$\infty$-Seq2 and B-Seq is duplicated.
If the language can throw exceptions we need even more (inductive)
rules in order to correctly propagate such abrupt termination, which
further increases duplication:\footnote{It is also possible to
  propagate exceptions automatically when they occur in tail-positions
  in big-step rules. This would make rule B-Exc-Seq2 redundant and
  eliminate some of the redundancy for the sequential composition
  construct. It would, however, require extra restrictions in the
  standard inductive rule for sequential composition. The duplication
  problem occurs in any case for constructs with more than two
  premises.}
\[
  \inference{
    ( c_1,\sigma ) => \<exc>(v)
  }{
    ( c_1; c_2, \sigma ) =>
    \<exc>(v)
  }[B-Exc-Seq1]
  \qquad
  \inference{
    ( c_1,\sigma ) => \sigma'
    &
    ( c_2,\sigma' ) => \<exc>(v)
  }{
    ( c_1; c_2, \sigma ) => \<exc>(v)
  }[B-Exc-Seq2]
\]
The semantics of sequential composition is now given by five rules
with eight premises, where two of these premises are duplicates (i.e.,
the first premise in B-$\infty$-Seq2, and B-Exc-Seq2).

Chargu\'{e}raud \cite{Chargueraud2013prettybigstep} introduced the
novel \emph{pretty-big-step} style of big-step SOS. Pretty-big-step
rules avoid the duplication problem by breaking big-step rules into
smaller rules such that each rule fully evaluates a single sub-term
and then continues the evaluation. The pretty-big-step rules introduce
an intermediate expression constructor, $\<seq2>$, and use
\emph{outcomes} $o$ to range over either convergence at a store
$\sigma$ ($\<conv>\ \sigma$), divergence ($\<div>$), or abrupt
termination ($\<exc>(v)$):
\begin{gather*}
  \inference{
    ( c_1,\sigma ) =v o_1
    &
    ( \mathsf{seq2}\ o_1\ c_2, \sigma2 ) =v o
  }{
    ( c_1; c_2, \sigma ) =v
    o
  }[P-Seq1]
  \quad
  \inference{
    ( c, \sigma ) =v o
  }{
    ( \mathsf{seq2}\ (\mathsf{conv}\ \sigma)\ c, \sigma_0 ) =v
    o
  }[P-Seq2]
  \quad
  \inference{
    \<abort>(o)
  }{
    ( \mathsf{seq2}\ o\ c_2, \sigma) =v
    o
  }[P-Seq-Abort]
\end{gather*}
Following Chargu\'{e}raud, these rules have a \emph{dual}
interpretation: inductive and coinductive. They use an $\<abort>$
predicate to propagate either exceptions or abrupt termination. This
predicate is specified once-and-for-all and not on a
construct-by-construct basis:
\[
  \inference{
  }{
    \<abort>(\<div>)
  }[Abort-Div]
  \qquad
  \inference{
  }{
    \<abort>(\<exc>(v))
  }[Abort-Exc]
\]
Using pretty-big-step, the semantics for sequential composition has
three rules with three premises, and two generic rules for the abort
predicate. It avoids duplication by breaking the original inductive
big-step rule B-Seq into smaller rules. But it also increases the
number of rules compared with the original inductive big-step rule
B-Seq. For semantics with rules with more premises but without abrupt
termination, pretty-big-step semantics sometimes \emph{increases} the
number of rules and premises compared with traditional inductive
big-step rules.

In this paper we reuse the idea from pretty-big-step semantics of
interpreting the same set of rules both inductively and coinductively,
and adopt and adapt the technique (due to Klin
\cite[p.~216]{Mosses2004MSOS}) for modular specification of abrupt
termination in small-step Modular SOS
\cite{Mosses2004MSOS}.
The idea is to make program states and result states record an
additional \emph{status flag} (ranged over by $\delta$) which
indicates that the current state is either convergent ($||d$),
divergent ($||u$), or abruptly terminated (`$\<exc>(v)$'):
\[
  \inference{
    ( c_1,\sigma, ||d ) => \sigma', \delta
    &
    ( c_2,\sigma',\delta ) => \sigma'', \delta'
  }{
    ( c_1; c_2, \sigma, ||d ) =>
    \sigma'',\delta'
  }[F-Seq]
\]
Here, derivations continue only so long as they are in a convergent
state, indicated by $||d$. In order to propagate divergence or abrupt
termination, we use pretty-big-step inspired abort rules:
\[
  \inference{
  }{
    (c, \sigma, ||u) => \sigma', ||u
  }[F-Div]
  \qquad
  \inference{
  }{
    (c, \sigma, \<exc>(v)) => \sigma', \<exc>(v)
  }[F-Exc]
\]
The abort rules say that all parts of the state except for the status
flag are computationally irrelevant, hence the use of the free
variables $\sigma'$ in F-Div and F-Exc. Using these status flags, we
can prove that terms diverge similarly to using either traditional
big-step divergence rules or pretty-big-step semantics. For example,
using pretty-big-step, the proposition $(c, \sigma) =v \<div>$ is
coinductively proved when $(c, \sigma)$ has an infinite derivation
tree; similarly, using the $||u$ status flag, the proposition, for
any $\sigma'$, $(c, \sigma, ||d ) => \sigma', ||u$ is coinductively
proved when $(c, \sigma, ||d )$ has an infinite derivation tree.

We call this style of rules \emph{flag-based big-step
  semantics}. Flag-based big-step semantics:
\begin{itemize}
\item supports propagating exceptions without the intermediate
  expression forms found in pretty-big-step rules (such as
  $\mathsf{seq2}$);\footnote{There is, however, nothing to prevent us
    from using the flag-based approach for propagating divergence in
    pretty-big-step rules. Section~\ref{sec:beyond-while} discusses
    why this could be attractive for certain applications.}
\item uses fewer rules and premises than both the traditional and the
  pretty-big-step approach;
\item supports reasoning about possibly-infinite computations on a par
  with traditional big-step approaches as well as small-step
  semantics; and
\item eliminates the big-step duplication problem for diverging and
  abruptly terminating computations.
\end{itemize}

The rest of this paper is structured as follows. We first introduce a
simple While-language and recall how possibly-diverging computations
in small-step semantics are traditionally expressed
(Sect.~\ref{sec:lang-small-step}). Next, we recall traditional
approaches to representing possibly-diverging computations in big-step
semantics, and how to prove equivalence between semantics for these
approaches (Sect.~\ref{sec:big-step-and-variants}). Thus equipped,
we make the following contributions:
\begin{itemize}
\item We present a novel approach to representing divergence and
  abrupt termination (Sect.~\ref{sec:gen-div}) which alleviates the
  duplication problem in big-step semantics. For all examples
  considered by the authors, including applicative and imperative
  languages, our approach straightforwardly allows expressing
  divergence and abrupt termination in a way that does not involve
  introducing or modifying rules. Our approach uses fewer rules and
  premises than both small-step semantics and pretty-big-step
  semantics.
\item We consider how the approach scales to more interesting language
  features (Sect.~\ref{sec:beyond-while}), including
  non-deterministic interactive input. A problem in this connection is
  that the traditional proof method for relating diverging
  computations in small-step and big-step SOS works only for
  deterministic semantics.  We provide a generalised proof method
  which suffices to relate small-step and big-step semantics with
  non-deterministic interactive input.
\item
  The explicit use of flags in our approach makes the rules somewhat
  tedious to read and write. We suggest leaving the flag arguments
  implicit in almost all rules
  (Sect.~\ref{sec:implicit-flag-based-divergence}).
  The way we specify this is similar to Implicitly-Modular SOS (I-MSOS)
  \cite{Mosses2009implicitpropagation}.
\end{itemize}
Our experiments show that flag-based big-step SOS with divergence and
abrupt termination uses fewer rules than previous approaches. The
conciseness comes at the cost of states recording irrelevant
information (such as the structure of the store) in abruptly
terminated or divergent result states. We discuss and compare previous
approaches in Sect.~\ref{sec:related-work} and conclude in
Sect.~\ref{sec:conclusion}.

\section{The While-Language and its Small-Step Semantics}
\label{sec:lang-small-step}
We use a simple While-language as a running example. Its abstract
syntax is:
\begin{align*}
  \mathit{Var} \ni x &::= \mathrm{x} \mid \mathrm{y} \mid \ldots & \text{Variables}\\
  \mathbb{N} \ni n &::= 0 \mid 1 \mid \ldots & \text{Natural numbers}\\
  \mathit{Cmd} \ni c &::= \mathsf{skip} \mid \mathsf{alloc}\ x \mid x
                       := e \mid c ; c \mid \textsf{if}\ e\ c\ c \mid
                       \textsf{while}\; e\; c &\text{Commands} \\
  \mathit{Val} \ni v &::= \mathsf{null} \mid n & \text{Values} \\
  \mathit{Expr} \ni e &::= v \mid x \mid
                        e \oplus e & \text{Expressions} \\
  \oplus &\in \{ +, -, * \} & \arraycolsep=0pt
  \begin{array}{r}
    \text{Binary operations}\\
    \text{on natural numbers}
  \end{array}
\end{align*}
Here, $\mathsf{null}$ is a special value used for uninitialised
locations, and is assumed not to occur in source programs. Let stores
$\sigma \in \mathit{Var} -fin-> \mathit{Val}$ be finite maps from
variables to values.  We use $\sigma(x)$ to denote the value to which
variable $x$ is mapped (if any) in the store $\sigma$. The notation
$\sigma[x |-> v]$ denotes the update of store $\sigma$ with value $v$
at variable $x$. We use $\mathrm{dom}(\sigma)$ to denote the domain of
a map, and $\{x_1 |-> v_1, x_2 |-> v_2,\ldots\}$ to denote the map
from $x_1$ to $v_1$, $x_2$ to $v_2$, etc.  We write $\oplus(n_1,n_2)$
for the result of applying the primitive binary operation $\oplus$ to
$n_1$ and $n_2$. The remainder of this section introduces a mixed
big-step and small-step semantics for this language, and introduces
conventions.

\subsection{Big-Step Expression Evaluation Relation}
Expressions in our language do not affect the store and cannot
diverge, although they can fail to produce a value. Big-step rules for
evaluating expressions and the signature of the evaluation relation
are given in Fig.~\ref{fig:big-exp}.\footnote{Following Reynolds
  \cite{Reynolds1972definitionalinterpreters}, this relation is
  \emph{trivial}, and expression evaluation could have been given in
  terms of an auxiliary function instead. It is useful to model it as
  a relation for the purpose of our approach to abrupt termination. We
  discuss this further in Sect.\ref{sec:beyond-while}.} A judgment
$(e, \sigma) =E=> v$ says that evaluating $e$ in the store $\sigma$
results in value $v$, and does not affect the store.

\begin{figure}
  \hfill
  \framebox{$(e, \sigma) =E=> v$}
  \begin{gather*}
    \inference{}{(v,\sigma) =E=> v}[E-Val]
    \qquad
    \inference{
      x \in \mathrm{dom}(\sigma)
    }{
      ( x,\sigma) =E=> \sigma(x)
    }[E-Var]
    \qquad
    \inference{
      (e_1,\sigma) =E=> n_1
      &
      (e_2,\sigma) =E=> n_2
    }{
      ( e_1 \oplus e_2,\sigma) =E=>
      \oplus(n_1,n_2)
    }[E-Bop]
  \end{gather*}
  \caption{Big-step semantics for expressions}
  \label{fig:big-exp}
\end{figure}

\subsection{Small-Step Command Transition Relation}
Commands have side-effects and can diverge. Figure~\ref{fig:small-cmd}
defines a small-step transition relation for commands, using the
previously defined big-step semantics for expressions. The transition
relation is defined for states consisting of pairs of a command $c$
and a store $\sigma$. A judgment $(c, \sigma) -> (c', \sigma')$
asserts the possibility of a transition from the state $(c, \sigma)$
to the state $(c', \sigma')$.

\begin{figure}
  \hfill
  \framebox{$(c, \sigma) -> (c', \sigma')$}
  \begin{gather*}
    \inference{
      x \not\in \mathrm{dom}(\sigma)
    }{
      ( \mathsf{alloc}\ x, \sigma ) -> ( \mathsf{skip}, \sigma[x |->
       \mathsf{null} ] )
    }[S-Alloc]
    \qquad
    \inference{
      x \in \mathrm{dom}(\sigma)
      &
      (e,\sigma) =E=> v
    }{
      ( x := e, \sigma ) -> ( \mathsf{skip},
      \sigma[x |-> v] )
    }[S-Assign]
    \\
    \inference{
      ( c_1,\sigma ) -> ( c_1', \sigma' )
    }{
      ( c_1; c_2, \sigma ) ->
      ( c_1'; c_2,\sigma' )
    }[S-Seq]
    \qquad
    \inference{
    }{
      ( \mathsf{skip}; c_2, \sigma ) ->
      ( c_2, \sigma )
    }[S-SeqSkip]
    \\
    \inference{
      (e,\sigma) =E=> v & v\neq 0
    }{
      ( \textsf{if}\ e\ c_1\ c_2, \sigma ) ->
      ( c_1,\sigma )
    }[S-If]
    \qquad
    \inference{
      ( e,\sigma) =E=> 0
    }{
      ( \textsf{if}\ e\ c_1\ c_2, \sigma ) ->
      ( c_2, \sigma )
    }[S-IfZ]
    \\
    \inference{
      (e,\sigma) =E=> v & v\neq 0
    }{
      ( \textsf{while}\ e\ c, \sigma ) ->
      ( c; \textsf{while}\ e\ c, \sigma )
    }[S-While]
    \qquad
    \inference{
      (e,\sigma) =E=> 0
    }{
      ( \textsf{while}\ e\ c, \sigma ) ->
      ( \mathsf{skip}, \sigma )
    }[S-WhileZ]
  \end{gather*}
  \caption{Small-step semantics for commands}
  \label{fig:small-cmd}
\end{figure}

\subsection{Finite Computations}
\label{ssec:finite-computations}
In small-step semantics, computations are given by sequences of
transitions. The $->*$ relation is the reflexive-transitive closure of
the transition relation for commands, which contains the set of all
computations with finite transition sequences:
\begin{gather}
  \tag*{\framebox{$(c, \sigma) ->* (c', \sigma')\vphantom{-inf->}$}}\\
  \inference{}{
    ( c, \sigma ) ->* ( c, \sigma )
  }[Refl$*$]
  \qquad
  \inference{
    ( c, \sigma ) -> ( c', \sigma' )
    &
    ( c', \sigma' ) ->* ( c'', \sigma'' )
  }{
    ( c, \sigma ) ->* ( c'', \sigma'' )
  }[Trans$*$]
  \tag*{}
\end{gather}
The following factorial function is an example of a program with a
finite sequence of transitions for any natural number $n$:
\[
\begin{array}{lcll}
  \mathit{fac}\ n &\equiv& \mathsf{alloc}\ \mathrm{c};\ \mathrm{c} := n;\\
                  &      & \mathsf{alloc}\ \mathrm{r};\ \mathrm{r} := 1;\\
                  &      & \mathsf{while}\ \mathrm{c}\
                                            (\mathrm{r} :=
                                            (\mathrm{r} * \mathrm{c}); \\
                  &      &                  \qquad\qquad\, \ \mathrm{c}:=
                                            (\mathrm{c} - 1))
\end{array}
\]
Here, `$\mathit{fac}\ n \equiv \ldots$' defines a function
$\mathit{fac}$ that, given a natural number $n$, produces a program
calculating the factorial of $n$. Using `$\cdot$' to denote the empty
map, we can use $->*$ to calculate:
\[
(\mathit{fac}\ 4,\cdot) ->* (\mathsf{skip}, \{\mathrm{c} |-> 0,
\mathrm{r} |-> 24\})
\]
This calculation is performed by constructing the finite derivation
tree whose conclusion is the judgment above.

\subsection{Infinite Computations}
\label{ssec:infinite-computations}
The following rule for $-inf->$ is \emph{coinductively} defined, and
can be used to reason about \emph{infinite sequences} of transitions:
\begin{gather}
  \tag*{\framebox{$(c, \sigma) -inf->$}}\\
  \inference[\textsf{co}]{
    ( c, \sigma ) ->
    ( c', \sigma' )
    &
    ( c', \sigma' ) -inf->
  }{
    ( c, \sigma ) -inf->
  }[Trans$\infty$]
  \tag*{}
\end{gather}
Here and throughout this article, we use `\textsf{co}' on the left of
rules to indicate that the relation is coinductively defined by that
set of rules.\footnote{This notation for coinductively defined
  relations is a variation of Cousot and Cousot's
  \cite{Cousot1992inductivedefinitions} notation for distinguishing
  inductively and coinductively (or \emph{positively} and
  \emph{negatively}) defined relations.} We assume that the reader is
familiar with the basics of coinduction (for introductions see, e.g.,
\cite{Leroy2009coinductivebigstep,Pierce2002TAPL,Sangiorgi2012anintroduction}).

Usually, coinductively defined rules describe both finite and infinite
derivation trees. However, since there are no axioms for $-inf->$, and
since the rule Trans$\infty$ cannot be used to construct finite
derivation trees, the relation contains exactly the set of all states
with infinite sequences of transitions.

Using the coinduction proof principle allows us to construct infinite
derivation trees. For example, we can prove that $\mathsf{while}\ 1\
\mathsf{skip}$ diverges by using $(\mathsf{while}\ 1\
\mathsf{skip},\cdot) -inf->$ as our \emph{coinduction hypothesis}. In
the following derivation tree, that hypothesis is applied at the point
in the derivation tree marked CIH. This constructs an infinite branch
of the derivation tree:
\[
\arraycolsep=0pt
\inference{
  \inference{
    (1, \cdot) =E=> 1 & 1 \neq 0
  }{
    \begin{array}{l}
      (\mathsf{while}\ 1\ \mathsf{skip}, \cdot) ->\\
      \quad(\mathsf{skip}; \mathsf{while}\ 1\ \mathsf{skip}, \cdot)
    \end{array}
  }
  &
  \inference{
    \vcenter{\inference{
      }{
        {\begin{array}{l}
            (\mathsf{skip}; \mathsf{while}\ 1\ \mathsf{skip}, \cdot) ->\\
            \quad(\mathsf{while}\ 1\ \mathsf{skip}, \cdot)
          \end{array}}
      }}
    &
    \inference{
      \vdots
    }{
      (\mathsf{while}\ 1\ \mathsf{skip}, \cdot) -inf->
    }[CIH]
    \!\!\!\!\!\!\!\!\!\!\!\!\!\!
  }{
    (\mathsf{skip}; \mathsf{while}\ 1\ \mathsf{skip}, \cdot) -inf->
  }
}{
  (\mathsf{while}\ 1\ \mathsf{skip}, \cdot) -inf->
}
\]

There also exists an infinite derivation tree whose conclusion is:
\[
(\mathsf{alloc}\ \mathrm{x}; \mathrm{x}:=0; \mathsf{while}\ 1\
(\mathrm{x}:= \mathrm{x} + 1),\cdot) -inf->
\]
Proving this is slightly more involved: after two applications of
Trans$\infty$, the goal to prove is:
\[
  (\mathsf{while}\ 1\ (\mathrm{x}:= \mathrm{x} + 1),\{\mathrm{x} |->
  0\}) -inf->
\]
We might try to use this goal as the coinduction hypothesis. But after
three additional applications of Trans$\infty$, we get a goal with a
store $\{ \mathrm{x} |-> 1 \}$ that does not match this coinduction
hypothesis:
\[
  (\mathsf{while}\ 1\ (\mathrm{x}:= \mathrm{x} + 1), \{\mathrm{x}
  |-> 1\}) -inf->
\]
The problem here is that the store changes with each step.  Instead,
we first prove the following straightforward lemma by coinduction:
\[
\text{for any }n,\
(\mathsf{while}\ 1\ (\mathrm{x} := \mathrm{x} + 1), \{\mathrm{x} |->
n\}) -inf->
\]
Now, by four applications of Trans$\infty$ in the original proof
statement, we get a goal that matches our lemma, which completes the
proof.

\subsection{Proof Conventions}
The formal results we prove in this article about our example language
are formalised in Coq and are available online at:
\url{http://www.plancomps.org/flag-based-big-step/}.

Coq is based on the Calculus of Constructions
\cite{Coquand1988thecalculus,Pierce2004ATTAPL}, which embodies a
variant of constructive logic. Working within this framework,
classical proof arguments, such as the law of excluded middle, are not
provable for arbitrary propositions. In spite of embodying a
constructive logic, Coq allows us to assert classical arguments as
axioms, which are known to be consistent \cite{Seldin1997ontheproof}
with Coq's logic. Some of the proofs in this article rely on the law
of excluded middle. For the reader concerned with implementing proofs
in Coq, or other proof assistants or logics based on constructive
reasoning, we follow the convention of Leroy and Grall
\cite{Leroy2009coinductivebigstep} and explicitly mark proofs that
rely on the law of excluded middle ``\emph{(classical)}''.

\section{Big-Step Semantics and Their Variants}
\label{sec:big-step-and-variants}
We recall different variants of big-step semantics from the
literature, and illustrate their use on the While-language defined in
the previous section (always extending the big-step semantics for
expressions defined in Fig.~\ref{fig:big-exp}).

\subsection{Inductive Big-Step Semantics}
\label{ssec:ind-big-step}
The big-step rules in Fig.~\ref{fig:big-step} inductively define a
big-step relation, where judgments of the form
$(c, \sigma) =B=> \sigma'$ assert that a command $c$ evaluated in
store $\sigma$ terminates with a final store $\sigma'$. This
corresponds to arriving at a state $(\<skip>,\sigma)$ by analogy with
the small-step semantics in Sect.~\ref{sec:lang-small-step}.
\begin{figure}
  \hfill
  \framebox{$(c, \sigma) =B=> \sigma'$}
  \begin{gather*}
    \inference{
    }{
      (\mathsf{skip}, \sigma) =B=> \sigma
    }[B-Skip]
    \qquad
    \inference{
      x \not\in \mathrm{dom}(\sigma)
    }{
      ( \mathsf{alloc}\ x, \sigma ) =B=>
      \sigma[x |-> \mathsf{null} ]
    }[B-Alloc]
    \\
    \inference{
      x\in\mathrm{dom}(\sigma)
      &
      (e,\sigma) =E=> v
    }{
      ( x := e, \sigma ) =B=>
      \sigma[x |-> v]
    }[B-Assign]
    \qquad
    \inference{
      ( c_1,\sigma ) =B=> \sigma'
      &
      ( c_2,\sigma' ) =B=> \sigma''
    }{
      ( c_1; c_2, \sigma ) =B=>
      \sigma''
    }[B-Seq]
    \\
    \inference{
      (e,\sigma) =E=> v & v \neq 0
      &
      ( c_1, \sigma ) =B=> \sigma'
    }{
      ( \textsf{if}\ e\ c_1\ c_2, \sigma ) =B=>
      \sigma'
    }[B-If]
    \qquad
    \inference{
      (e,\sigma) =E=> 0
      &
      ( c_2, \sigma ) =B=> \sigma'
    }{
      ( \textsf{if}\ e\ c_1\ c_2, \sigma ) =B=>
      \sigma'
    }[B-IfZ]
    \\
    \inference{
      (e,\sigma) =E=> v & v \neq 0
      &
      ( c, \sigma ) =B=> \sigma'
      \\
      ( \textsf{while}\ e\ c, \sigma' ) =B=> \sigma''
    }{
      ( \textsf{while}\ e\ c, \sigma ) =B=>
      \sigma''
    }[B-While]
    \qquad
    \inference{
      (e,\sigma) =E=> 0
    }{
      ( \textsf{while}\ e\ c, \sigma ) =B=>
      \sigma
    }[B-WhileZ]
  \end{gather*}
  \caption{Big-step semantics for commands}
  \label{fig:big-step}
\end{figure}
Being inductively defined, the relation does not contain diverging
programs.
\begin{example}
  $\neg \ \exists \sigma . \ (\mathsf{while}\ 1\ \mathsf{skip},
  \cdot) =B=> \sigma$
  \begin{proof}
    We prove that
    $\exists \sigma . \ (\mathsf{while}\ 1\ \mathsf{skip}, \cdot) =B=>
    \sigma$
    implies falsity. The proof proceeds by eliminating the existential
    in the premise, and by induction on the structure of $=B=>$.
  \end{proof}
\end{example}
We can, however, construct a finite derivation tree for our factorial
program:
\[
(\mathit{fac}\ 4,\cdot) =B=> \{\mathrm{c} |-> 0, \mathrm{r} |-> 24\}
\]
The following theorem proves the correspondence between inductive
big-step derivation trees and derivation trees for sequences of
small-step transitions for the While-language:
\begin{theorem}
  $(c, \sigma) ->* (\mathsf{skip}, \sigma')$ iff $(c, \sigma) =B=>
  \sigma'$.
  \begin{proof}
    The small-to-big direction follows by induction on $->*$ using
    Lemma~\ref{lem:bigsmall2big}. The big-to-small direction follows
    by induction on $=B=>$, using the transitivity of $->*$ and
    Lemma~\ref{lem:star-seq-cong}.
  \end{proof}
  \label{lem:small-big-conv}
\end{theorem}

\begin{lemma}
  If $(c, \sigma) -> (c', \sigma')$ and $(c', \sigma') =B=> \sigma''$
  then $(c,\sigma) =B=> \sigma''$.
  \begin{proof}
    Straightforward proof by induction on $->$.
  \end{proof}
  \label{lem:bigsmall2big}
\end{lemma}

\begin{lemma}
  $(c_1, \sigma) ->* (c_1', \sigma')$ implies $(c_1;c_2,\sigma) ->*
  (c_1'; c_2, \sigma')$
  \begin{proof}
    Straightforward proof by induction on $->*$.
  \end{proof}
  \label{lem:star-seq-cong}
\end{lemma}

\subsection{Coinductive Big-Step Divergence Predicate}
\label{ssec:coind-div-pred}
\begin{figure}
\hfill \framebox{$(c, \sigma) =inf=>$}
  \begin{gather*}
    \inference[\textsf{co}]{
      (c_1, \sigma) =inf=>
    }{
      (c_1; c_2, \sigma) =inf=>
    }[D-Seq1]
    \qquad
    \inference[\textsf{co}]{
      (c_1, \sigma) =B=> \sigma'
      &
      (c_2,\sigma') =inf=>
    }{
      (c_1; c_2, \sigma) =inf=>
    }[D-Seq2]
    \\
    \inference[\textsf{co}]{
      (e, \sigma) =E=> v & v\neq 0
      &
      (c_1, \sigma) =inf=>
    }{
      (\textsf{if}\ e\ c_1\ c_2, \sigma) =inf=>
    }[D-If]
    \qquad
    \inference[\textsf{co}]{
      (e, \sigma) =E=> 0
      &
      (c_2, \sigma) =inf=>
    }{
      (\textsf{if}\ e\ c_1\ c_2,\sigma) =inf=>
    }[D-IfZ]
    \\
    \inference[\textsf{co}]{
      (e, \sigma) =E=> v & v\neq 0
      &
      (c, \sigma) =inf=>
    }{
      (\textsf{while}\ e\ c, \sigma) =inf=>
    }[D-WhileBody]
    \qquad
    \inference[\textsf{co}]{
      (e, \sigma) =E=> v & v\neq 0
      &
      (c, \sigma) =B=> \sigma'
      \\
      (\textsf{while}\ e\ c, \sigma') =inf=>
    }{
      (\textsf{while}\ e\ c, \sigma) =inf=>
    }[D-While]
  \end{gather*}
  \caption{Big-step semantics for diverging commands (extending
    Figure~\ref{fig:big-step})}
  \label{fig:div-pred}
\end{figure}%
The rules in Fig.~\ref{fig:div-pred} coinductively define a big-step
divergence predicate. Like `$-inf->$', the `$=inf=>$' predicate has no
axioms, so the rules describe exactly the set of all diverging
computations. Divergence arises in a single branch of a derivation
tree, while other branches may converge. Recalling our program from
Sect.~\ref{sec:lang-small-step}:
\[
  \mathsf{alloc}\ \mathrm{x}; \mathrm{x}:=0; \mathsf{while}\ 1\
  (\mathrm{x}:= \mathrm{x} + 1)
\]
Here, $\mathsf{alloc}\ \mathrm{x}$ and $\mathrm{x}:=0$ converge, but
the $\mathsf{while}$ command diverges. The big-step divergence
predicate uses the standard inductive big-step relation for converging
branches. Consequently, we need the rules in \emph{both}
Figures~\ref{fig:big-step}~and~\ref{fig:div-pred} to express
divergence. This increases the number of rules for each construct, and
leads to duplication of premises between the rules. For example, the
set of finite and infinite derivation trees whose conclusion is a
sequential composition requires three rules: B-Seq, D-Seq1, and
D-Seq2. Here, the premise $(c_1, \sigma) =B=> \sigma'$ is duplicated
between the rules, giving rise to the so-called duplication problem
with big-step semantics.  Theorem~\ref{thm:smallinfiffbiginf} proves
the correspondence between $-inf->$ and $=inf=>$.
\begin{theorem}
  $(c, \sigma) -inf->$ iff $(c,\sigma) =inf=>$
  \label{lem:small-big-div}
  \begin{proof}[Proof (classical)]
    The small-to-big direction follows by coinduction using
    Lemma~\ref{lem:classicsmall} and Lemma~\ref{lem:infseqcong} (which
    relies on classical arguments) for case analysis. The other
    direction follows by coinduction and
    Lemma~\ref{lem:biginf2biginfsmall}.
  \end{proof}
  \label{thm:smallinfiffbiginf}
\end{theorem}

\begin{lemma}
  Either $(\exists \sigma'.\ (c,\sigma) ->* (\mathsf{skip},\sigma'))$
  or $(c,\sigma) -inf->$.
  \begin{proof}[Proof (classical)]
    By the law of excluded middle and classical reasoning.
  \end{proof}
  \label{lem:classicsmall}
\end{lemma}

\begin{lemma}
  If $(c_1, \sigma) ->* (c_1', \sigma')$ and $(c_1;c_2,\sigma)
  -inf->$, then $(c_1';c_2,\sigma') -inf->$.
  \begin{proof}
    Straightforward proof by induction on $->*$, using the fact that
    $->$ is deterministic.
  \end{proof}
  \label{lem:infseqcong}
\end{lemma}

\begin{lemma}
  If $(c,\sigma) =inf=>$ then $\exists c', \sigma'.\ \left((c,\sigma)
    -> (c',\sigma') \ \wedge \ (c',\sigma') =inf=> \right)$.
  \begin{proof}
    Straightforward proof by structural induction on the command $c$,
    using Theorem~\ref{lem:small-big-conv} in the sequential
    composition case.
  \end{proof}
  \label{lem:biginf2biginfsmall}
\end{lemma}

\subsection{Pretty-Big-Step Semantics}
\label{ssec:pretty-big-step-semantics}
The idea behind pretty-big-step semantics is to break big-step rules
into intermediate rules, so that each rule fully evaluates a single
sub-term and then continues the evaluation. Continuing evaluation may
involve either further computation, or propagation of divergence or
abrupt termination that arose during evaluation of a
sub-term. Following Chargu\'{e}raud
\cite{Chargueraud2013prettybigstep}, we introduce so-called
\emph{semantic constructors} for commands and \emph{outcomes} for
indicating convergence or divergence:
\begin{gather*}
  \mathit{SemCmd} \ni C ::=  c \mid \mathsf{assign2}\ x\ v \mid
  \mathsf{seq2}\ o\ c \mid \textsf{if2}\ v\ c\ c
  \mid \textsf{while2}\ v\ e\ c \mid \textsf{while3}\ o\ e\ c
  \\
  \mathit{Outcome} \ni o ::= \mathsf{conv}\ \sigma \mid \mathsf{div}
\end{gather*}
The added constructors are used to distinguish whether evaluation
should continue.

For example, the following rules define sequential composition for a
pretty-big-step relation with the signature `$(C, \sigma) =v o$',
using the semantic constructor $\mathsf{seq2}$:
\[
\inference{
  (c_1, \sigma) =v o_1
  &
  (\mathsf{seq2}\ o_1\ c_2, \sigma) =v o
}{
  (c_1; c_2, \sigma) =v o
}[P-Seq1]
\qquad
\inference{
  (c_2,\sigma) =v o
}{
  (\mathsf{seq2}\ (\mathsf{conv}\ \sigma)\ c_2,\sigma_0) =v o
}[P-Seq2]
\]
Each of these two rules is a pretty-big-step rule: reading the rules
in a bottom-up manner, P-Seq1 evaluates a single sub-term $c_1$, plugs
the result of evaluation into the semantic constructor
$\mathsf{seq2}$, and evaluates that term. The rule P-Seq2 in turn
checks that the result of evaluating the first sub-term was
convergence with some store $\sigma$, and goes on to evaluate $c_2$
under that store. An additional so-called \emph{abort rule} is
required for propagating divergence if it occurs in the first branch
of a sequential composition:
\[
\inference{
}{
  (\mathsf{seq2}\ \mathsf{div}\ c_2, \sigma) =v \mathsf{div}
}[P-Seq-Abort]
\]
Such abort rules are required for all semantic constructors. As
Chargu\'{e}raud \cite{Chargueraud2013prettybigstep} remarks, these are
tedious both to read and write, but are straightforward to generate
automatically.

The pretty-big-step rules in Fig.~\ref{fig:pretty-big} have a
\emph{dual} interpretation: such rules define two separate relations,
one \emph{inductive}, the other \emph{coinductive}. We use
`\textsf{du}' on the left of rules to indicate relations with dual
interpretations. We use $=v$ to refer to the inductive interpretation,
and $=co=v$ to refer to the coinductive interpretation. We refer to
the union of the two interpretations of the relation defined by a set
of pretty-big-step rules by $=du=v$. Crucially, these relations are
based on the same set of rules. In practice, this means that the
coinductively defined relation subsumes the inductively defined
relation, as shown by Lemma~\ref{lem:pbig2copbig}.
\begin{lemma}
  If $(C, \sigma) =v o$ then $(C,\sigma) =co=v o$.
  \begin{proof}
    Straightforward proof by induction on $=v$.
  \end{proof}
  \label{lem:pbig2copbig}
\end{lemma}
However, the coinductive interpretation is less useful for proving
properties about converging programs, since converging and diverging
programs cannot be syntactically distinguished in the coinductive
interpretation. For example, we can prove that $\mathsf{while}\ 1\
\mathsf{skip}$ coevaluates to anything:
\begin{example}
  For any $o$, $(\mathsf{while}\ 1\ \mathsf{skip}, \cdot) =co=v o$.
  \begin{proof}
    Straightforward proof by coinduction.
  \end{proof}
  \label{ex:pbs-while1}
\end{example}

\begin{figure}
  \hfill \framebox{$(C, \sigma) =v o$}
  \begin{gather*}
    \inference[\textsf{du}]{
    }{
      (\mathsf{skip}, \sigma) =v \mathsf{conv}\ \sigma
    }[P-Skip]
    \qquad
    \inference[\textsf{du}]{
      x \not\in \mathrm{dom}(\sigma)
    }{
      ( \mathsf{alloc}\ x, \sigma ) =v
      ( \mathsf{conv}\ \sigma[x |-> \mathsf{null} ])
    }[P-Alloc]
    \\
    \inference[\textsf{du}]{
      (e,\sigma) =E=> v
      &
      (\mathsf{assign2}\ x\ v,\sigma) =v o
    }{
      (x := e, \sigma ) =v o
    }[P-Assign1]
    \qquad
    \inference[\textsf{du}]{
      x \in \mathrm{dom}(\sigma)
    }{
      (\mathsf{assign2}\ x\ v, \sigma ) =v
      \mathsf{conv}\ \sigma[x |-> v]
    }[P-Assign2]
    \\
    \inference[\textsf{du}]{
      ( c_1,\sigma ) =v o_1
      &
      ( \mathsf{seq2}\ o_1\ c_2, \sigma ) =v o
    }{
      ( c_1; c_2, \sigma ) =v o
    }[P-Seq1]
    \qquad
    \inference[\textsf{du}]{
      ( c, \sigma ) =v o
    }{
      ( \mathsf{seq2}\ (\mathsf{conv}\ \sigma)\ c, \sigma_0 ) =v
      o
    }[P-Seq2]
    \\
    \inference[\textsf{du}]{
      (e,\sigma) =E=> v
      \\
      ( \textsf{if2}\ v\ c_1\ c_2, \sigma ) =v o
    }{
      ( \textsf{if}\ e\ c_1\ c_2, \sigma ) =v o
    }[P-If]
    \qquad
    \inference[\textsf{du}]{
      v \neq 0
      &
      ( c_1, \sigma ) =v o_1
    }{
      ( \textsf{if2}\ v\ c_1\ c_2, \sigma ) =v
      o_1
    }[P-If2]
    \qquad
    \inference[\textsf{du}]{
      ( c_2, \sigma ) =v o_2
    }{
      ( \textsf{if2}\ 0\ c_1\ c_2, \sigma ) =v o_2
    }[P-IfZ2]
    \\
    \inference[\textsf{du}]{
      (e,\sigma) =E=> v
      &
      ( \textsf{while2}\ v\ e\ c, \sigma) =v o
    }{
      ( \textsf{while}\ e\ c, \sigma ) =v o
    }[P-While]
    \qquad
    \inference[\textsf{du}]{
      v \neq 0
      \\
      ( c, \sigma ) =v o
      &
      ( \textsf{while3}\ o\ e\ c, \sigma ) =v o'
    }{
      ( \textsf{while2}\ v\ e\ c, \sigma) =v o'
    }[P-While2]
    \\
    \inference[\textsf{du}]{
    }{
      ( \textsf{while2}\ 0\ e\ c, \sigma) =v
      \mathsf{conv}\ \sigma
    }[P-WhileZ2]
    \qquad\!
    \inference[\textsf{du}]{
      ( \textsf{while}\ e\ c, \sigma ) =v o
    }{
      \textsf{while3}\ (\mathsf{conv}\ \sigma)\ e\ c, \sigma_0) =v
      o
    }[P-While3]
    \\
    \inference[\textsf{du}]{
    }{
      ( \mathsf{seq2}\ \mathsf{div}\ c_2, \sigma) =v
      \mathsf{div}
    }[P-Seq-Abort]
    \quad
    \inference[\textsf{du}]{
    }{
      ( \mathsf{while3}\ \mathsf{div}\ e\ c, \sigma) =v
      \mathsf{div}
    }[P-While-Abort]
  \end{gather*}
  \caption{Pretty-big-step semantics for commands}
  \label{fig:pretty-big}
\end{figure}

An important property of the rules in Fig.~\ref{fig:pretty-big} is
that divergence is only derivable under the coinductive
interpretation.
\begin{lemma}
  $(c,\sigma) =v o$ implies $o \neq \mathsf{div}$.
  \begin{proof}
    Straightforward proof by induction on $=v$.
  \end{proof}
  \label{lem:pbignotdiv}
\end{lemma}

Pretty-big-step semantics can be used to reason about terminating
programs on a par with traditional big-step relations, as shown by
Theorem~\ref{lem:big2pbig}.
\begin{theorem}
  $(c,\sigma) =B=> \sigma'$ iff $(c,\sigma) =v \mathsf{conv}\ \sigma'$.
  \begin{proof}
    Each direction is proved by straightforward induction on $=B=>$
    and $=v$ respectively. The $=v$-to-$=B=>$ direction uses
    Lemma~\ref{lem:pbignotdiv}.
  \end{proof}
  \label{lem:big2pbig}
\end{theorem}
Pretty-big-step semantics can be used to reason about diverging
programs on a par with traditional big-step divergence predicates, as
shown by Theorem~\ref{thm:biginf2pbigdiv}.
\begin{theorem}
  $(c,\sigma) =inf=>$ iff $(c,\sigma) =co=v \mathsf{div}$.
  \begin{proof}[Proof (classical)]
    Each direction is proved by coinduction. The $=inf=>$-to-$=co=v$
    direction uses Lemma~\ref{lem:big2pbig}. The other direction uses
    Lemma~\ref{lem:biginforbig} (which relies on classical arguments)
    and Lemma~\ref{lem:pbigcopbigdet}.
  \end{proof}
  \label{thm:biginf2pbigdiv}
\end{theorem}
\begin{lemma}
  If $(c,\sigma) =co=v o$ and $\neg \left( (c,\sigma) =v o\right)$
  then $(c,\sigma) =co=v \mathsf{div}$
  \begin{proof}[Proof (classical)]
    By coinduction, using Lemma~\ref{lem:pbig2copbig} and the law of
    excluded middle for case analysis on $(c,\sigma) =v (o,\sigma')$.
  \end{proof}
  \label{lem:biginforbig}
\end{lemma}
\begin{lemma}
  If $(c,\sigma) =v \mathsf{conv}\ \sigma'$ and $(c,\sigma) =co=v
  \mathsf{conv}\ \sigma''$ then $\sigma' = \sigma''$.
  \begin{proof}
    Straightforward proof by induction on $=v$.
  \end{proof}
  \label{lem:pbigcopbigdet}
\end{lemma}

Our pretty-big-step semantics uses 18 rules (including rules for
$=E=>$) with 16 premises (counting judgments about both $=E=>$ and
$=du=v$), none of which are duplicates. In contrast, the union of the
rules for inductive standard big-step rules in
Sect.~\ref{ssec:ind-big-step} and the divergence predicate in
Sect.~\ref{ssec:coind-div-pred} uses 17 rules with 25 premises of
which 6 are duplicates. Pretty-big-step semantics solves the
duplication problem, albeit at the cost of introducing 5 extra
semantic constructors and breaking the rules in the inductive
interpretation up into multiple rules, which in this case adds an
extra rule compared to the original big-step rules with duplication.

\section{Flag-Based Big-Step Semantics}
\label{sec:gen-div}
In this section we present a novel approach to representing divergence
and abrupt termination in big-step semantics, which does not require
introducing new relations, and relies on fewer, simpler rules for
propagation of divergence and/or abrupt termination. The approach can
be used to augment standard inductively defined rules to allow them to
express divergence on a par with standard divergence predicates or
pretty-big-step rules.

\subsection{The While-Language and its Flag-Based Big-Step Semantics}
\label{ssec:gen-div-big}
We show how augmenting the standard inductive big-step rules from
Figures~\ref{fig:big-exp}~and~\ref{fig:big-step} with \emph{status
  flags} allows us to express and reason about divergence on a par
with traditional big-step divergence predicates. Status flags indicate
either convergence ($||d$) or divergence ($||u$), ranged over by:
\[
  \mathit{Status} \ni \delta ::= \text{$||d$} \mid \text{$||u$}
\]
Threading status flags through the conclusion and premises of the
standard big-step rules in left-to-right order gives the rules in
Fig.~\ref{fig:big-div-state}. In all rules, the status flag is
threaded through rules such that the conclusion source always starts
in a $||d$
state.

Like pretty-big-step semantics, the rules in
Fig.~\ref{fig:big-div-state} have a \emph{dual} interpretation: both
inductive and coinductive. We use $=G=>$ to refer to the relation
given by the inductive interpretation of the rules, and $=coG=>$ to
refer to the relation given by the coinductive interpretation. When
needed, we refer to the union of these interpretations by $=duG=>$.

Figure~\ref{fig:big-div-state} threads status flags through the rules
for the $=E=>$ relation as well as the standard big-step relation for
commands, $=G=>$. But expressions cannot actually diverge. Our
motivation for threading the flag through the rules for expression
evaluation is that it anticipates future language extensions which may
permit expressions to diverge, such as allowing user-defined functions
to be called. It also allows us to correctly propagate divergence in
rules where evaluation of expressions does not necessarily occur as
the first premise in rules.

\begin{figure}
  \hfill
  \framebox{$(e, \sigma, \delta) =GE=> v, \delta'$}
  \begin{gather*}
    \inference[\<du>]{}{(v,\sigma, ||d) =GE=> v, ||d}[FE-Val]
    \qquad
    \inference[\<du>]{
      x \in \mathrm{dom}(\sigma)
    }{
      ( x,\sigma, ||d) =GE=> \sigma(x), ||d
    }[FE-Var]
    \qquad
    \inference[\<du>]{
      (e_1,\sigma, ||d) =GE=> n_1, \delta
      \\
      (e_2,\sigma, \delta) =GE=> n_2, \delta'
    }{
      ( e_1 \oplus e_2,\sigma, ||d) =GE=>
      \oplus(n_1, n_2), \delta'
    }[FE-Bop]
    \\
    \inference[\<du>]{
    }{
      (e,\sigma,||u) =GE=> v, ||u
    }[FE-Div]
  \end{gather*}
  \hfill
  \framebox{$(c, \sigma, \delta) =G=> \sigma', \delta'$}
  \begin{gather*}
    \inference[\textsf{du}]{
    }{
      (\mathsf{skip}, \sigma, ||d) =G=> \sigma, ||d
    }[F-Skip]
    \qquad
    \inference[\textsf{du}]{
      x \not\in \mathrm{dom}(\sigma)
    }{
      ( \mathsf{alloc}\ x, \sigma, ||d ) =G=>
      \sigma[x |-> \mathsf{null} ], ||d
    }[F-Alloc]
    \\
    \inference[\textsf{du}]{
      x \in\mathrm{dom}(\sigma)
      &
      (e,\sigma, ||d) =GE=> v, \delta
    }{
      ( x := e, \sigma, ||d ) =G=>
      \sigma[x |-> v ], \delta
    }[F-Assign]
    \qquad
    \inference[\textsf{du}]{
      ( c_1,\sigma,||d ) =G=> \sigma', \delta
      &
      ( c_2,\sigma',\delta ) =G=> \sigma'', \delta'
    }{
      ( c_1; c_2, \sigma, ||d) =G=>
      \sigma'', \delta'
    }[F-Seq]
    \\
    \inference[\textsf{du}]{
      v \neq 0
      \\
      (e,\sigma,||d) =GE=> v, \delta
      &
      ( c_1, \sigma, \delta) =G=> \sigma', \delta'
    }{
      ( \textsf{if}\ e\ c_1\ c_2, \sigma,||d ) =G=>
      \sigma',\delta'
    }[F-If]
    \qquad
    \inference[\textsf{du}]{
      ( e,\sigma,||d) =GE=> 0, \delta
      &
      ( c_2, \sigma, \delta ) =G=> \sigma', \delta'
    }{
      ( \textsf{if}\ e\ c_1\ c_2, \sigma, ||d ) =G=>
      \sigma',\delta'
    }[F-IfZ]
    \\
    \inference[\textsf{du}]{
      (e,\sigma,||d) =GE=> v,\delta
      &
      v \neq 0
      \\
      ( c, \sigma, \delta ) =G=> \sigma', \delta'
      &
      ( \textsf{while}\ e\ c, \sigma', \delta' ) =G=> \sigma'',\delta''
    }{
      ( \textsf{while}\ e\ c, \sigma, ||d ) =G=>
      \sigma'',\delta''
    }[F-While]
    \qquad
    \inference[\textsf{du}]{
      (e,\sigma,||d) =GE=> 0,\delta
    }{
      ( \textsf{while}\ e\ c, \sigma, ||d ) =G=>
      \sigma,\delta
    }[F-WhileZ]
    \\
    \inference[\textsf{du}]{
    }{
      (c,\sigma,||u) =G=> \sigma', ||u
    }[F-Div]
  \end{gather*}
  \caption{Flag-based big-step semantics for commands and expressions
    with divergence}
  \label{fig:big-div-state}
\end{figure}

How do these rules support reasoning about divergence?  In the F-Seq
rule in Fig.~\ref{fig:big-div-state}, the first premise may diverge
to produce $||u$ in place of $\delta$ in the first premise. If this is
the case, any subsequent computation is irrelevant. To inhibit
subsequent computation we introduce \emph{divergence rules} FE-Div and
F-Div, also in Fig.~\ref{fig:big-div-state}. The divergence rules
serve a similar purpose to abort rules in pretty-big-step semantics:
they propagate divergence as it arises and inhibit further evaluation.

A technical curiosity of the divergence rule F-Div is that it allows
evaluation to return an \emph{arbitrary} store. In other words, states
record and propagate irrelevant information. This is a somewhat
unusual way of propagating semantic information in big-step semantics;
however, the intuition behind it follows how divergence and abrupt
termination are traditionally propagated in big-step SOS. Recall that
big-step and pretty-big-step rules discard the structure of the
current program term, and only record that divergence occurs. Witness,
for example, the big-step and pretty-big-step rules for propagating
exceptions from Sect.~\ref{sec:introduction}:
\[
  \inference{
    ( c_1,\sigma ) =inf=>
  }{
    ( c_1; c_2, \sigma ) =>
    \sigma''
  }[B-$\infty$-Seq1]
  \qquad
  \inference{
    \<abort>(o)
  }{
    ( \mathsf{seq2}\ o\ c_2, \sigma) =v
    o
  }[P-Seq-Abort]
\]
These rules ``forget'' the structure of whatever other effects a
diverging computation produces.
Similarly, flag-based big-step semantics retains the relevant
structure of configurations, but allows us to choose arbitrary values
for the irrelevant parts: for converging computations, all parts of
the configuration are relevant, whereas for divergent or abruptly
terminated configurations, only the fact that we are abruptly
terminating is important.

A key property of flag-based divergence is that the conclusions of
rules always start in a state with the convergent flag $||d$. It
follows that, under an inductive interpretation, we cannot construct
derivations that result in a divergent state $||u$.
Lemma~\ref{lem:gbignotbig} proves that we cannot use the inductively
defined relation to prove divergence.
\begin{lemma}
  $(c,\sigma,||d) =G=> \sigma',\delta$ implies $\delta \neq \text{$||u$}$.
  \begin{proof}
    Straightforward proof by induction on $=G=>$.
  \end{proof}
  \label{lem:gbignotbig}
\end{lemma}

Theorem~\ref{lem:big-gd-conv} proves that adding status flags and the
divergence rule does not change the inductive meaning of the standard
inductive big-step relation.
\begin{theorem}
  $(c, \sigma) =B=> \sigma'$ iff $(c, \sigma, ||d) =G=> \sigma',||d$.
  \begin{proof}
    The $=B=>$-to-$=G=>$ follows by straightforward induction. The
    $=G=>$-to-$=B=>$ direction follows by straightforward induction
    and Lemma~\ref{lem:gbignotbig}.
  \end{proof}
  \label{lem:big-gd-conv}
\end{theorem}
As for the coinductive interpretation of $=G=>$,
Theorem~\ref{lem:big-gd-div} proves that adding status flags allows us
to prove divergence on a par with traditional big-step divergence
predicates.
\begin{theorem}
  For any $\sigma'$, $(c,\sigma) =inf=>$ iff
  $(c,\sigma,||d) =coG=> \sigma',||u$.
  \begin{proof}[Proof (classical)]
    The $=inf=>$-to-$=coG=>$ direction follows by straightforward
    coinduction, using Theorem~\ref{lem:big-gd-conv}. The other
    direction follows from Lemma~\ref{lem:gbigdivorgbig},
    Lemma~\ref{lem:gbig2cogbig}, and the law of excluded middle for
    case analysis on whether branches converge or not.
  \end{proof}
  \label{lem:big-gd-div}
\end{theorem}
\begin{lemma}
  If $(c,\sigma,||d) =coG=> \sigma', \delta$ and
  $\neg \left( (c,\sigma,||d) =G=> \sigma', \delta \right)$ then
  $(c,\sigma,||d) =coG=> \sigma', ||u$.
  \begin{proof}[Proof (classical)]
    By coinduction and the law of excluded middle for case analysis on
    $=G=>$.
  \end{proof}
  \label{lem:gbigdivorgbig}
\end{lemma}

The relationship between the inductive and coinductive interpretation
of the rules for $=G=>$ is similar to the relationships observed in
connection with pretty-big-step semantics. The coinductive
interpretation also subsumes the inductive interpretation, as proved
in Lemma~\ref{lem:gbig2cogbig}.
\begin{lemma}
  If $(c,\sigma,||d) =G=> \sigma',||d$ then $(c,\sigma,||d) =coG=>
  \sigma',||d$.
  \begin{proof}
    Straightforward proof by induction on $=G=>$.
  \end{proof}
  \label{lem:gbig2cogbig}
\end{lemma}

Theorem~\ref{lem:big-gd-conv} proves that we can choose \emph{any}
store $\sigma'$ when constructing a proof of divergence for some $c$
and $\sigma$. Lemma~\ref{lem:store-irrelevant} summarises this
observation, by proving that the choice of $\sigma'$ is, in fact,
irrelevant.

\begin{lemma}
  For any $c$ and any store $\sigma$, $(\forall \sigma'. (c,\sigma,||d)
  =coG=> (\sigma',||d))$ iff $(\exists \sigma'.\ (c, \sigma, ||d)
  =coG=> (\sigma',||d))$.
  \begin{proof}
    The $\forall$-to-$\exists$ direction is trivial. The other
    direction follows by straightforward coinduction.
  \end{proof}
  \label{lem:store-irrelevant}
\end{lemma}

As with pretty-big-step semantics
(Sect.~\ref{ssec:pretty-big-step-semantics}), the coinductive
interpretation is less useful for proving properties about converging
programs, since converging and diverging programs cannot be
distinguished in the coinductive interpretation. For example, we can
prove that $\mathsf{while}\ 1\ \mathsf{skip}$ coevaluates to anything:
\begin{example}
  For any $\sigma'$ and $\delta$,
  $(\mathsf{while}\ 1\ \mathsf{skip}, \cdot, ||d) =coG=>
  \sigma',\delta$.
  \begin{proof}
    Straightforward proof by coinduction.
  \end{proof}
  \label{lem:while1skipdiv}
\end{example}

Comparing flag-based divergence (Fig.~\ref{fig:big-div-state}) with
pretty-big-step (Fig.~\ref{fig:pretty-big}), we see that our rules
contain no duplication, and use just 13 rules with 13 premises,
whereas pretty-big-step semantics uses 18 rules with 16
premises. While we use fewer rules than pretty-big-step, and
introducing divergence does not introduce duplicate premises, the
flag-based big-step rules in Fig.~\ref{fig:big-div-state} actually
do contain some duplicate premises in the rules F-If, F-IfZ, F-While,
and F-WhileZ: each of these rules evaluate the same expression
$e$. This duplication could have been avoided by introducing a
constructor for intermediate expressions for `$\<if>$' and
`$\<while>$', similar to pretty-big-step.

Flag-based big-step SOS is equivalent to but more concise than
traditional big-step and pretty-big-step. For the simple
While-language, it is no more involved to work with flag-based
divergence than with traditional approaches, as illustrated by
examples in our Coq proofs available online at:
\url{http://www.plancomps.org/flag-based-big-step/}.

\subsection{Divergence Rules are Necessary}
\label{ssec:properties-big-gen-div}
From Example~\ref{lem:while1skipdiv} we know that some diverging
programs result in a $||d$ state. But do we really need the $||u$ flag
and divergence rule? Here, we answer this question
affirmatively. Consider the following program:
\[
(\textsf{while}\ 1\ \textsf{skip}); \mathsf{alloc}\ \mathrm{x};
\mathrm{x}:= \mathrm{x} + 0
\]
Proving that this program diverges depends crucially on F-Div allowing
us to propagate divergence. The while-command diverges whereas the
last sub-commands of the program contain a stuck computation: the
variable $\mathrm{x}$ has the $\mathsf{null}$ value when it is
dereferenced, so this program will attempt to add $\mathsf{null}$ and
$0$, which is meaningless. The derivation trees we can construct must
use the F-Div rule as follows:
\[
\inference{
  {\arraycolsep=0pt
   \begin{array}{c}
     \vphantom{gT} \\
     (\textsf{while}\ 1\ \textsf{skip},\cdot,||d) =coG=>
     (\cdot,||u)
   \end{array}}
  &
  \inference{}{
    (\mathsf{alloc}\ \mathrm{x};
    \mathrm{x} := \mathrm{x} + 0, \cdot,||u) =coG=> (\cdot, ||u)
  }[F-Div]\!\!\!\!\!\!\!\!\!\!\!\!\!\!\!\!\!
}{
  ((\textsf{while}\ 1\ \textsf{skip}); \mathsf{alloc}\ \mathrm{x};
  \mathrm{x} := \mathrm{x} + 0, \cdot,||d) =coG=>
  (\cdot, ||u)
}[F-Seq]
\]

\begin{example}
  For any $\sigma$, $((\mathsf{while}\ 1\ \mathsf{skip});
  \mathsf{alloc}\ \mathrm{x}; \mathrm{x} := \mathrm{x} + 0,
  \cdot,||d) =co=> (\sigma, ||u)$. In contrast,\\
  $\neg\ \exists \sigma .\ ((\mathsf{while}\ 1\ \mathsf{skip});
    \mathsf{alloc}\ \mathrm{x}; \mathrm{x} := \mathrm{x} + 0,
    \cdot,||d) =co=> (\sigma, ||d)$
  \label{lem:while-assign-not-arbitrary}
\end{example}

Leroy and Grall \cite{Leroy2009coinductivebigstep} observed a similar
point about the coinductive interpretation of the rules for the
$\lambda$-calculus with constants, i.e., that it is non-deterministic,
and that it does not contain computations that diverge and then get
stuck, nor computations that diverge towards infinite values. Here, we
have shown (Theorem~\ref{lem:big-gd-conv}~and~\ref{lem:big-gd-div})
that coevaluation of flag-based big-step semantics is equally
expressive as traditional divergence predicates and, transitively (by
Theorem~\ref{lem:small-big-conv}~and~\ref{thm:smallinfiffbiginf}),
small-step semantics.

\section{Beyond the While-Language}
\label{sec:beyond-while}
Flag-based big-step semantics supports reasoning about divergence on a
par with small-step semantics for the simple While-language. In this
section we illustrate that the approach also scales to deal with other
language features, such as exceptions and non-deterministic input. To
this end, we present a novel proof method for relating small-step and
big-step semantics with sources of non-determinism. Finally, we
discuss potential pitfalls and limitations.

\subsection{Non-deterministic input}
\label{ssec:nondet}
Previous approaches to relating big-step and small-step SOS focus
mainly on relating finite computations
\cite{Nipkow2014concretesemantics,Pierce2013SF,Ciobaca2013fromsmallstep}. The
main counter-example appears to be Leroy and Grall's study of
coinductive big-step semantics \cite{Leroy2009coinductivebigstep},
which considers a deterministic language. Their proof (which is
analogous to Theorem~\ref{thm:smallinfiffbiginf} in
Sect.~\ref{ssec:coind-div-pred} of this article) for relating
diverging computations in small-step and big-step semantics relies
crucially on determinism. We consider the extension of our language
with an expression form for non-deterministic interactive input, and
prove that the extension preserves the equivalence of small-step
semantics and flag-based big-step semantics. Our proof provides a
novel approach to relating small-step and big-step SOS for divergent
computations involving non-determinism at the leaves of derivation
trees.

Consider the extension of our language with the following expression
form which models interactive input:
\[
  \mathit{Expr} \ni e ::= \ldots \mid \<input>
\]
Its single expression evaluation rule is:
\[
  \inference{
  }{
    (\<input>,\sigma,||d) =GE=> v, ||d
  }
\]
Adding this construct clearly makes our language non-deterministic,
since the value $v$ to the right of $=GE=>$ is arbitrary. If we extend
expression evaluation for small-step correspondingly, is the big-step
flag-based semantics still equivalent to the small-step semantics?

The answer to this question should intuitively be ``yes'': we have not
changed the semantics of commands, so most of the proofs of the
properties about the relationship between small-step and big-step
semantics from Sect.~\ref{sec:lang-small-step} carry over unchanged,
since they rely on the expression evaluation relation being equivalent
between the two semantics.

But the proof of Theorem~\ref{thm:smallinfiffbiginf} crucially relies
on Lemma~\ref{lem:infseqcong}, which in turn holds only for
deterministic transition relations and expression evaluation
relations. The violated property is the following:
\begin{quote}
  If $(c_1, \sigma) ->* (c_1', \sigma')$ and $(c_1;c_2,\sigma)
  -inf->$, then $(c_1';c_2,\sigma') -inf->$.
\end{quote}
Here, if $->$ is non-deterministic, it may be that $c_1$ can make a
sequence of transition steps to a $c_1'$ which is stuck, whereas there
may exist some alternative sequence of transition steps for $c_1$ such
that $c_1;c_2$ diverges.
\begin{example}
  Consider the program state where $\mathrm{x}$ is an unbound
  variable: {\rm
  \[
    (\overbrace{\<if>\ \<input>\ (\mathrm{x} := 0)\ \<skip>}^{c_1};
    \overbrace{\<while>\ 1\ \<skip>}^{c_2},\cdot)
  \]
}%
If {\rm $\<input>$} returns a non-zero value, evaluation gets stuck,
since the command `$\mathrm{x}:=0$' is meaningless in the empty store
(`$\cdot$'). Otherwise, evaluation diverges. Thus it holds that
$(c_1,\cdot) ->* (\mathrm{x}:= 0,\cdot)$ and $(c_1;c_2,\cdot) -inf->$,
but not $(\mathrm{x}:= 0; c_2,\cdot) -inf->$.
\end{example}
What we can prove instead about possibly-terminating computations that
rely on sequential composition is the following:
\begin{lemma}
  If $(c_1;c_2,\sigma) -inf->$ and {\rm
    $\neg (\exists \sigma'.\ (c_1,\sigma) ->* (\<skip>,\sigma'))$},
  then it must be the case that $(c_1,\sigma) -inf->$.
  \begin{proof}
    The proof is by coinduction, using the goal as coinduction
    hypothesis. The proof follows by inversion on the first
    hypothesis, from which we derive that either {\rm
      $c_1 = \<skip>$}, which leads to a contradiction, or there
    exists a configuration $(c_1',\sigma_1)$ for which
    $(c_1,\sigma) -> (c_1',\sigma_1)$. By the second hypothesis, it
    must also hold that there is no $\sigma'$ such that {\rm
      $(c_1',\sigma_1) ->* (\<skip>, \sigma')$}. From these facts, the
    goal follows by applying Trans$\infty$, the small-step rule S-Seq,
    and the coinduction hypothesis.
  \end{proof}
  \label{lem:infseqcong-alt}
\end{lemma}
Using Lemma~\ref{lem:infseqcong-alt}, we can relate infinite sequences
of small-step transitions to infinite flag-based big-step derivations
as follows.
\begin{lemma}
  For any $\sigma'$, if $(c,\sigma) -inf->$ then
  $(c,\sigma,||d) =coG=> \sigma',||u$.
  \begin{proof}[Proof (classical)]
    The proof is by guarded coinduction, using the goal as the
    coinduction hypothesis. The critical cases are those for
    sequential composition and `$\<while>$': these cases use the law
    of excluded middle for case analysis on whether $c$ converges or
    not. If it does, the goal follows by Lemma~\ref{lem:gbig2cogbig}
    (which carries over unchanged). If it does not, the goal follows
    by Lemma~\ref{lem:infseqcong-alt} above.
  \end{proof}
\end{lemma}
The proof of the other direction, i.e., relating infinite big-step
derivations to infinite sequences of small-step transitions is proved
using Lemma~\ref{lem:biginf2biginfsmall-sb}:
\begin{theorem}
  $(c, \sigma) -inf->$ iff for any $\sigma'$,
  $(c,\sigma,||d) =coG=> \sigma',||u$.
  \begin{proof}[Proof (classical)]
    The small-to-big direction is proved in
    Lemma~\ref{lem:infseqcong-alt} above. The other direction follows
    by coinduction and Lemma~\ref{lem:biginf2biginfsmall-sb} below.
  \end{proof}
  \label{thm:smallinfiffbiginf-sb}
\end{theorem}

\begin{lemma}
  For any $\sigma''$, if $(c,\sigma) =coG=> \sigma'',||u$, then
  $\exists c', \sigma'.\ \left((c,\sigma) -> (c',\sigma') \ \wedge \
    (c',\sigma') =coG=> \sigma'',||u \right)$.
  \begin{proof}
    The proof carries over from Lemma~\ref{lem:biginf2biginfsmall},
    and relies on Theorem~\ref{thm:small-big-conv-sb} below.
  \end{proof}
  \label{lem:biginf2biginfsmall-sb}
\end{lemma}

\begin{theorem}
  $(c, \sigma) ->* (\mathsf{skip}, \sigma')$ iff $(c, \sigma) =B=>
  \sigma'$.
  \begin{proof}
    The proof straightforwardly carries over from
    Theorem~\ref{lem:small-big-conv}.
  \end{proof}
  \label{thm:small-big-conv-sb}
\end{theorem}

\subsection{Exceptions}
\label{ssec:exceptions}
We extend our language with exceptions. We add exceptions to our
language by augmenting the syntactic sort for status flags:
\[
  \mathit{Status} \ni \delta ::= \ldots \mid \<exc>(v, \sigma)
\]
We let exceptions record both a thrown value and the state of the
store when the exception occurs. We also augment the syntactic sort
for commands with a `$\<throw>\ v$' construct for throwing a value $v$
as an exception, and a `$\<catch>\ c\ c$' construct for handling
exceptions:
\[
  \mathit{Cmd} \ni c ::= \ldots \mid \<throw>\ v \mid \<catch>\ c\ c
\]
The rules for these constructs are given by a single rule for throwing
an exception, and two rules for propagating exceptions for command and
expression evaluation:
\begin{gather*}
  \inference{
  }{
    (\<throw>\ v,\sigma,||d) =G=> \sigma',\<exc>(v, \sigma)
  }[F-Throw]
  \\
  \inference{
  }{
    (c,\sigma,\<exc>(v, \sigma_0)) =G=> \sigma',\<exc>(v, \sigma_0)
  }[F-Exc]
  \qquad
  \inference{
  }{
    (e,\sigma,\<exc>(v,\sigma_0)) =GE=> v',\<exc>(v,\sigma_0)
  }[FE-Exc]
\end{gather*}
Using these rules, exceptions are propagated similarly to divergence.

The following rules specify the semantics of `$\<catch>\ c\ c$':
\begin{gather*}
  \inference{
    (c_1, \sigma, ||d) =G=> \sigma',\delta
    &
    \delta \neq \<exc>(\_, \_)
  }{
    (\<catch>\ c_1\ c_2, \sigma, ||d) =G=> \sigma', \delta
  }[F-Catch]
  \qquad
  \inference{
    (c_1, \sigma, ||d) =G=> \sigma',\<exc>(v, \sigma_0)
    &
    (c_2, \sigma_0, ||d) =G=> \sigma'', \delta
  }{
    (\<catch>\ c_1\ c_2, \sigma, ||d) =G=> \sigma'',\delta
  }[F-Catch-Some]
\end{gather*}
Here, F-Catch handles the case where no exception arises during
evaluation of $c_1$. F-Catch-Some detects that an exception has
occurred and proceeds to evaluate the handler, $c_2$, in the store
recorded in the exception.\footnote{The handler $c_2$ here discards
  the value of the exception. It is straightforward to give flag-based
  semantics for handlers with patterns that match thrown exception
  values, e.g., by adapting rules from
  \cite[Sect.~3.3]{Churchill2015reusable} or
  \cite[Fig.~2]{Churchill2013modularbisimulationtheory}.}

\subsection{The Necessity of Choosing Arbitrary Stores}
\label{ssec:the-need-for-choosing}
The F-Throw and F-Exc rules above admit arbitrary $\sigma'$s on the
right-hand side of the arrow. Admitting arbitrary stores to be
propagated in connection with divergence was motivated in part by the
fact that divergent computations are non-deterministic anyway (see,
e.g., Example~\ref{lem:while1skipdiv}). But computations that throw
exceptions are guaranteed to converge, so do we really need it here?

For the simple language considered here, we do not strictly need to
relate exceptional states to arbitrary other states. But consider a
variant of the While-language where exceptions can arise during
expression evaluation, where expression evaluation may affect the
store, and where we allow variables to be passed around as first-class
values similarly to references in Standard ML
\cite{Milner1997thedefinition}. The signature of expression evaluation
and the assignment rule in such a language could be:
\begin{gather*}
  \tag*{\framebox{$(e, \sigma, \delta) =GE=> v, \sigma, \delta'$}}
  \\
  \inference{
    (e_1,\sigma,||d) =GE=> x,\sigma',\delta
    &
    (e_2,\sigma',\delta) =GE=> v,\sigma'',\delta'
    &
    x\in\mathrm{dom}(\sigma'')
  }{
    (e_1 := e_2,\sigma,||d) =GE=> v,\sigma''[x |-> v],\delta'
  }[F-RefAsgn]
\end{gather*}
Given the rule and language described above, if $e_1$ or $e_2$
abruptly terminates the rule still insists that
$x\in\mathrm{dom}(\sigma'')$. Thus, it becomes essential that abrupt
termination allows us to return an arbitrary store, such that we can
synthesise a store $\sigma''$ for which
$x\in\mathrm{dom}(\sigma'')$. If we did not, evaluation might get
stuck instead of propagating abrupt termination.

A further consequence of a rule like F-RefAsgn above is that the
property summarised in Lemma~\ref{lem:store-irrelevant} no longer
holds: we can no longer prove that a program that diverges has an
arbitrary store. For some applications this is unimportant; for
others, this state of affairs is unfortunate. For example, a proof
assistant like Coq does not support coinductive reasoning about
existentially quantified goals in a satisfactory manner: Coq's support
for coinductive proofs is limited to guarded coinduction, which makes
many otherwise simple proofs unnecessarily involved. For example,
proving a goal of the following form is not possible by guarded
coinduction alone:
\[
  \text{If }P\text{ then } \exists \sigma'.\ (c,\sigma,||d) =co=> (\sigma', ||u)
\]
where $P$ is some proposition.

While the need to synthesise semantic information that is not the
result of any actual computation may be unattractive for certain
applications, flag-based big-step rules do allow abrupt termination
and divergence to be propagated correctly. Furthermore, the flag-based
approach is also applicable to the more flexible pretty-big-step
style: applying a flag-based propagation strategy to pretty-big-step
rules would alleviate the need for abort-rules for each semantic
expression constructor, since propagation is handled by a single
flag-based abort rule instead.\footnote{This approach was explored in
  previous work \cite[Sect.~3.1]{BachPoulsen2014derivingpretty} by
  the authors.}

\section{Implicit Flag-Based Divergence}
\label{sec:implicit-flag-based-divergence}

As illustrated in the previous sections, the flag-based approach can reduce the number of rules and premises required for specifying divergence or abrupt termination in big-step semantics. However, explicit threading of the flag arguments $\delta$ and $\delta'$ through the evaluation formulae in all rules is somewhat tedious, and might discourage adoption of the approach.

In this section, we introduce new notation for signatures. Using such signatures, the flag arguments can be left implicit in almost all rules. For instance, we can specify the While language with the signatures specified below simply by adding the divergence rules FE-Div and F-Div from Fig.~\ref{fig:big-div-state} to the original big-step rules given in Figs.~\ref{fig:big-exp} and~\ref{fig:big-step}.

Recall the original signatures used for the big-step semantics of While:
\[
  \framebox{$(e, \sigma) =E=> v$}\qquad
  \framebox{$(c, \sigma) =B=> \sigma'$}
\]
Our new signatures for the flag-based semantics of While are as follows:
\[
\framebox{$( e, \sigma \hibox{$, \delta {:}{-}{||d} $}) =GE=>
      v \hibox{$, \delta'$}$}
\qquad
\framebox{$( c, \sigma \hibox{$, \delta {:}{-}{||d} $}) =G=>
      \sigma' \hibox{$, \delta'$}$}
\]
The \hibox{highlighting} of the arguments $\delta$ and $\delta'$ specifies formally that they can either be omitted or  made explicit (uniformly) when the evaluation relation concerned is used in a rule. The notation `$\delta {:}{-}{||d}$' indicates that $||d$ is the \emph{default} value of the flag $\delta$.\footnote
{`$\delta {:}{-}{||d}$' is reminiscent of Prolog: it indicates that $||d$ is the condition for $\delta$ to allow evaluation to proceed normally.}
Rules written using just the non-highlighted parts of the signatures abbreviate flag-based rules as follows:
\begin{itemize}
\item
When a rule without explicit flags has no evaluation premises, it abbreviates the flag-based rule where the flags of both the source and target of the conclusion are $||d$.
For example,
\[
    \inference{
    }{
      (\mathsf{skip}, \sigma) =G=> \sigma
    }[F-Skip]
\quad\text{abbreviates}\quad
    \inference[\textsf{du}]{
    }{
      (\mathsf{skip}, \sigma, ||d) =G=> \sigma, ||d
    }[F-Skip] .
\]
\item
When a rule without explicit flags has $n$ evaluation premises, it abbreviates the flag-based rule where the flags of the source of the first evaluation premise and of the source of the conclusion are $||d$; the flags of the target of evaluation premise $i$ and of the source of evaluation premise $i+1$ are $\delta_i$ (for $1 \leq i < n$); and the flags of the target of evaluation premise $n$ and the conclusion are $\delta'$.
For example,
\[
    \inference{
      ( c_1,\sigma) =G=> \sigma'
      &
      ( c_2,\sigma') =G=> \sigma''
    }{
      ( c_1; c_2, \sigma) =G=>
      \sigma''
    }[F-Seq]
\quad\!\text{abbreviates}\quad\!
    \inference[\textsf{du}]{
      ( c_1,\sigma,||d ) =G=> \sigma', \delta_1
      &
      ( c_2,\sigma',\delta_1 ) =G=> \sigma'', \delta'
    }{
      ( c_1; c_2, \sigma, ||d) =G=>
      \sigma'', \delta'
    }[F-Seq] .
\]
\end{itemize}

Replacing the signatures in Figs.~\ref{fig:big-exp} and~\ref{fig:big-step} by those given above, the big-step rules there abbreviate all the flag-based rules given in Fig.~\ref{fig:big-div-state} (up to renaming of flag variables), so all that is needed is to add the two explicit divergence rules (FE-Div and F-Div).

\emph{The Definition of Standard ML} \cite{Milner1997thedefinition} introduced the technique of letting arguments of evaluation relations remain implicit in big-step rules, calling it ``the store convention''. I-MSOS \cite{Mosses2009implicitpropagation} proposed the use of highlighting to specify formally which arguments can be omitted, and defined I-MSOS by translation to MSOS \cite{Mosses2004MSOS}, but the notation provided there does not support the intended use of the default $||d$ in flag-based big-step rules.
It should however be possible to generalise I-MSOS to support flags, and thereby allow further arguments to be omitted. For instance, the I-MSOS signature
$\hibox{$($} \,c\, \hibox{$\vphantom{(}, \sigma )$} =G=>
      ( \hibox{$\vphantom{(} \sigma'$} )$
allows the $\sigma$ and $\sigma'$ arguments to be omitted as follows:
\[
    \inference{
    }{
      \mathsf{skip} =G=> (~)
    }[F-Skip]
\qquad
    \inference{
      c_1 =G=> (~)
      &
      c_2 =G=> (~)
    }{
      c_1; c_2 =G=> (~)
    }[F-Seq]
\]
The highlighting in the following signature could therefore specify that the above rules abbreviate flag-based MSOS rules:
\[
\framebox{$\hibox{$($} \,c\, \hibox{$\vphantom{(}, \sigma, \delta {:}{-}{||d} )$} =G=>
      ( \hibox{$\sigma' , \delta'$} )$}
\]
Further development of the details of implicit flag-based I-MSOS is out of the scope of this paper, and left to future work.

\section{Related Work}
\label{sec:related-work}
Several papers have explored how to represent divergence in big-step
semantics. Leroy and Grall \cite{Leroy2009coinductivebigstep} survey
different approaches to representing divergence in coinductive
big-step semantics, including divergence predicates, trace-based
semantics, and taking the coinductive interpretation of standard
big-step rules. They conclude that traditional divergence predicates
are the most well-behaved, but increase the size of specifications by
around 40\%. The trace-based semantics of Leroy and Grall relies on
concatenating infinite traces for accumulating the full trace of rules
with multiple premises. Nakata and Uustalu \cite{Nakata2009tracebased}
propose a more elegant approach to accumulating traces based on `peel'
rules. Their approach is closely related to the \emph{partiality
  monad}, introduced by Capretta, and used by several authors to give
functional representations of big-step semantics, including Danielsson
\cite{Danielsson2012operationalsemantics} and Abel and Chapman
\cite{Abel2014normalization} (who call it the \emph{delay monad}). In
the partiality monad, functions either return a finitely delayed
result or an infinite trace of delays. Pir\'{o}g and Gibbons
\cite{Pirog2014thecoinductive} study the category theoretic
foundations of the resumption monad. Related to the resumption monad
is the interactive I/O monad by Hancock and Setzer
\cite{Hancock2000interactiveprograms} for modeling the behaviour of
possibly-diverging interactive programs.

While it is possible to specify and reason about operational semantics
by means of the partiality monad, it relies on the ability to express
mixed recursive/corecursive functions in order to use it in proof
assistants. Agda \cite{Bove2009abrief} provides native support for
such function definitions, while Coq does not. Nakata and Uustalu
\cite{Nakata2009tracebased,Nakata2010resumptions} show that it is
possible to express and reason about a functional representation of a
trace-based semantics in Coq using a purely coinductive style. This
works well for the simple While-language, but preliminary experiments
suggest that such guarded coinductive functions can be subtle to
implement in Coq. Propagating divergence between premises in
trace-based big-step semantics is also slightly more involved than the
approach taken in pretty-big-step semantics and by us: propagating
divergence between premises in trace-based semantics entails a
coinductive proof for each premise (proving that the trace is
infinite). In contrast, propagation in both pretty-big-step semantics
and our approach relies on simple case analysis (i.e., via abort rules
or inspecting the status flag).

Trace-based semantics provides a strong foundation for constructively
reasoning about possibly-diverging programs, but unfortunately, as
discussed earlier, a modern proof assistant like Coq is not up to the
task of expressing and reasoning about these in a fully satisfactory
manner: Coq's support for coinductive proofs is limited to guarded
coinduction, which makes many otherwise simple proofs involving
existentials unnecessarily involved. For example, proving a goal of
the following form is not possible by guarded coinduction alone:
\[
  \text{If }P\text{ then } \exists \tau.\ (c,\sigma) =co=> \tau
\]
Here, $P$ is some proposition, and $\tau$ is a trace. Proving such
propositions can be done by manually constructing a witness function
for $\tau$, which is not always trivial. Chargu\'{e}raud
\cite{Chargueraud2013prettybigstep} directs a similar criticism at
Coq's lack of support for coinduction.

Hur et al.\ \cite{Hur2013thepower} present a novel means of doing
coinductive proofs. The idea is to use a \emph{parameterised greatest
  fixed-point} for proofs by coinduction. Their Coq library,
\emph{Paco}, provides an implementation of the method, which supports
proofs by coinduction using a more flexible notion of guardedness than
the naive syntactic guardedness check which Coq implements. This line
of work provides a promising direction for more tractable coinductive
proofs.

Looking beyond Coq, the support for features pertaining to coinduction
seems more progressive in the Agda proof assistant. Indeed, Danielsson
\cite{Danielsson2012operationalsemantics} leverages some of this
support by using mutually recursive/corecursive functions (which Coq
does not support) in his big-step functional operational semantics for
the $\lambda$-calculus. Also working in Agda, Abel and Chapman
\cite{Abel2014normalization} use \emph{sized types}
\cite{Abel2010miniagda} for their proofs by coinduction.

Nakata and Uustalu also give coinductive big-step semantics for
\emph{resumptions} \cite{Nakata2010resumptions} (denoting the external
behaviour of a communicating agent in concurrency theory
\cite{Milner1973processes}) as well as interleaving and concurrency
\cite{Uustalu2013coinductivebigstep}. These lines of work provide a
more general way of giving and reasoning about all possible outcomes
of a program than the committed-choice non-determinism that we
considered in Sect.~\ref{ssec:nondet}. It is, however, also somewhat
more involved to specify and work with.

Owens et al.\ \cite{Owens2016functionalbigstep} argue that
step-indexed semantics, i.e., using a counter which is decremented
with each recursive call, is an under-utilised technique for
representing big-step semantics with possible divergence. By
augmenting a big-step semantics with counters, it becomes possible to
represent it as a function in logic such that it is guaranteed to
terminate: when the clock runs out, a special ``time-out'' value is
produced. This permits reasoning about divergence, since the set of
diverging programs is exactly the set of programs for which there does
not exist a finitely-valued counter such that the program successfully
terminates without timing out. This provides an alternative that
avoids the duplication problem with big-step semantics but still
supports reasoning about diverging computations, albeit somewhat more
indirectly than representing infinite computations as infinite
derivation trees.

Moggi \cite{Moggi1991notionsofcomputation} suggested monads as a means
to obtain modularity in denotational semantics
\cite{Cenciarelli1993asyntactic}. In a similar vein, Modular SOS
\cite{Mosses2004MSOS} provides a means to obtain modularity in
operational semantics. The flag-based approach to divergence presented
here is a variant of the abrupt termination technique used in
\cite{Mosses2004MSOS}. That article represents abrupt termination as
emitted signals. In a small-step semantics, this enables the encoding
of abrupt termination by introducing a top-level handler that matches
on emitted signals: if the handler observes an emitted signal, the
program abruptly terminates. Exceptions as emitted signals could also
be used for expressing abrupt termination in big-step
semantics. However, this would entail wrapping each premise in a
handler, thereby cluttering rules. Subsequent work
\cite{BachPoulsen2014derivingpretty} observed that encoding abrupt
termination as a stateful flag instead scales better to big-step rules
by avoiding such explicit handlers. Flag-based divergence was used in
\cite{BachPoulsen2015imperativepolymorphism} for giving a semantics
for the untyped $\lambda$-calculus. The first author's thesis
\cite[Cpt.~4 and 5]{BachPoulsen2016xtss} describes a rule format for
which small-step and pretty-big-step rules are provably equivalent
using flag-based abrupt termination and divergence.

Our work differs from
\cite{BachPoulsen2014derivingpretty,BachPoulsen2015imperativepolymorphism,BachPoulsen2016xtss}
in several ways: unlike \cite{BachPoulsen2014derivingpretty} it
considers both inductive and coinductive big-step semantics and it is
based on SOS. In our work with Torrini
\cite{BachPoulsen2015imperativepolymorphism}, we proposed the
flag-based approach as a straightforward way of augmenting a semantics
such that it is useful for reasoning about divergence, but did not
provide proofs of the equivalence with traditional big-step semantics
nor small-step semantics. Another difference from this paper is that
both \cite{BachPoulsen2014derivingpretty,BachPoulsen2016xtss} mainly
considers pretty-big-step, whereas the flag-based big-step style
considered here has a more traditional big-step flavour.  The first
author's thesis \cite{BachPoulsen2016xtss} also uses the generalised
coinductive proof technique described in Sect.~\ref{ssec:nondet} for
relating divergent small-step and pretty-big-step semantics with
limited non-determinism. Whereas \cite{BachPoulsen2016xtss} applies
the technique to a variant of MSOS in pretty-big-step style, this
paper shows that the technique also scales to flag-based big-step SOS.

The question of which semantic style (small-step or big-step) is
better for proofs is a moot point. Big-step semantics are held to be
more convenient for certain proofs, such as compiler correctness
proofs. Leroy and Grall \cite{Leroy2009coinductivebigstep} cite
compiler correctness proofs as a main motivation for using coinductive
big-step semantics as opposed to small-step semantics: using
small-step semantics complicates the correctness proof. Indeed, Hutton
and Wright \cite{Hutton2007whatis} and Hutton and Bahr
\cite{Bahr2015calculating} also use big-step semantics for their
compiler correctness proofs. However, in \emph{CompCert}, Leroy
\cite{Leroy2006formalcertification} uses a small-step semantics and
sophisticated notions of bisimulation for its compiler correctness
proofs.

Wright and Felleisen introduce the syntactic approach to type safety
in \cite{Wright1994asyntactic}. They survey type safety proofs based
on denotational and big-step semantics, and conclude that small-step
semantics is a better fit for proving type safety by progress and
preservation lemmas. Harper and Stone \cite{Harper2000atypetheoretic}
direct a similar criticism at the big-step style. Big-step semantics
can, however, be used for strong type safety on a par with small-step
semantics. Leroy and Grall \cite{Leroy2009coinductivebigstep} show how
to coinductively prove progress using coinductive divergence
predicates, whereby type safety can be proved, provided one also
proves a big-step preservation lemma, which is usually
unproblematic. Another approach to big-step type safety consists in
making the big-step semantics \emph{total} by providing explicit error
rules for cases where the semantics goes wrong. Type safety is then
proved by showing that well-typed programs cannot go wrong. A
non-exhaustive list of examples that use explicit error rules
includes: Cousot's work on types as abstract interpretations
\cite{Cousot1997typesas}; Danielsson's work on operational semantics
using the partiality monad \cite{Danielsson2012operationalsemantics};
and Chargu\'{e}raud's work on pretty-big-step semantics
\cite{Chargueraud2013prettybigstep}, which provides a nice technique
for encoding explicit error rules more conveniently. A third option
for proving type safety using a big-step relation is to encode the
big-step semantics as a small-step abstract machine
\cite{Ager2003afunctional,Danvy2008ontheequivalence,Simmons2013alogical},
whereby the standard small-step type safety proof technique applies. A
preliminary study \cite{BachPoulsen2015imperativepolymorphism} of a
variant of Cousot's types as abstract interpretations suggests that
abstract interpretation can be used to prove big-step type safety
without the explicit error rules Cousot uses in his original
presentation \cite{Cousot1997typesas}.

\section{Conclusion}
\label{sec:conclusion}
We presented a novel approach to augmenting standard big-step
semantics to express divergence on a par with small-step and
traditional big-step semantics.  Our approach to representing
divergence uses fewer rules than existing approaches of similar
expressiveness (e.g., traditional big-step and pretty-big-step), and
the flag arguments of the evaluation relations can be generated
automatically.  We also considered how to extend our semantics with
interactive input, and provided a generalisation of the traditional
proof method for relating diverging small-step and big-step semantics.

Our experiments show that flag-based semantics provides a novel,
lightweight, and promising approach to concise big-step SOS
specification of programming languages involving divergence and abrupt
termination.

\paragraph{Acknowledgements} We thank the anonymous reviewers for
their constructive suggestions for improving this paper. Thanks also
to Neil Sculthorpe, Paolo Torrini, and Ulrich Berger for their helpful
comments on a previous version of this article.  This work was
supported by an EPSRC grant (EP/I032495/1) to Swansea University in
connection with the \emph{PLanCompS} project
(\texttt{\href{http://www.plancomps.org}{www.plancomps.org}}).

\section*{References}
\bibliography{jlamp2015}

\begin{thebibliography}{10}
\expandafter\ifx\csname url\endcsname\relax
  \def\url#1{\texttt{#1}}\fi
\expandafter\ifx\csname urlprefix\endcsname\relax\def\urlprefix{URL }\fi
\expandafter\ifx\csname href\endcsname\relax
  \def\href#1#2{#2} \def\path#1{#1}\fi

\bibitem{Plotkin2004astructuralapproach}
G.~D. Plotkin, A structural approach to operational semantics, J. Log. Algebr.
  Program. 60-61 (2004) 17--139.
\newblock \href {http://dx.doi.org/10.1016/j.jlap.2004.05.001}
  {\path{doi:10.1016/j.jlap.2004.05.001}}.

\bibitem{Kahn87naturalsemantics}
G.~Kahn, Natural semantics, in: STACS 87, Vol. 247 of LNCS, Springer, 1987, pp.
  22--39.
\newblock \href {http://dx.doi.org/10.1007/BFb0039592}
  {\path{doi:10.1007/BFb0039592}}.

\bibitem{Leroy2009coinductivebigstep}
X.~Leroy, H.~Grall, Coinductive big-step operational semantics, Inf. Comput.
  207~(2) (2009) 284--304.
\newblock \href {http://dx.doi.org/10.1016/j.ic.2007.12.004}
  {\path{doi:10.1016/j.ic.2007.12.004}}.

\bibitem{Nipkow2014concretesemantics}
T.~Nipkow, G.~Klein, Concrete Semantics: With {Isabelle/HOL}, Springer, 2014.

\bibitem{Hutton2007whatis}
G.~Hutton, J.~Wright, What is the meaning of these constant interruptions?, J.
  Functional Program. 17 (2007) 777--792.
\newblock \href {http://dx.doi.org/10.1017/S0956796807006363}
  {\path{doi:10.1017/S0956796807006363}}.

\bibitem{Danvy2004refocusingin}
O.~Danvy, L.~R. Nielsen, Refocusing in reduction semantics, BRICS Research
  Series RS-04-26, Dept. of Comp. Sci., Aarhus University (2004).

\bibitem{BachPoulsen2013generatingspecialized}
C.~Bach~Poulsen, P.~D. Mosses, Generating specialized interpreters for modular
  structural operational semantics, in: LOPSTR'13, Vol. 8901 of LNCS, Springer,
  2014, pp. 220--236.
\newblock \href {http://dx.doi.org/10.1007/978-3-319-14125-1\_13}
  {\path{doi:10.1007/978-3-319-14125-1\_13}}.

\bibitem{Chargueraud2013prettybigstep}
A.~Chargu{\'e}raud, Pretty-big-step semantics, in: ESOP'14, Vol. 7792 of LNCS,
  Springer, 2013, pp. 41--60.
\newblock \href {http://dx.doi.org/10.1007/978-3-642-37036-6_3}
  {\path{doi:10.1007/978-3-642-37036-6_3}}.

\bibitem{Mosses2004MSOS}
P.~D. Mosses, Modular structural operational semantics, J. Log. Algebr.
  Program. 60-61 (2004) 195--228.
\newblock \href {http://dx.doi.org/10.1016/j.jlap.2004.03.008}
  {\path{doi:10.1016/j.jlap.2004.03.008}}.

\bibitem{Mosses2009implicitpropagation}
P.~D. Mosses, M.~J. New, Implicit propagation in structural operational
  semantics, ENTCS 229~(4) (2009) 49--66.
\newblock \href {http://dx.doi.org/10.1016/j.entcs.2009.07.073}
  {\path{doi:10.1016/j.entcs.2009.07.073}}.

\bibitem{Reynolds1972definitionalinterpreters}
J.~C. Reynolds, Definitional interpreters for higher-order programming
  languages, in: ACM '72, ACM Annual Conference, ACM, New York, NY, USA, 1972,
  pp. 717--740.
\newblock \href {http://dx.doi.org/10.1145/800194.805852}
  {\path{doi:10.1145/800194.805852}}.

\bibitem{Cousot1992inductivedefinitions}
P.~Cousot, R.~Cousot, Inductive definitions, semantics and abstract
  interpretations, in: POPL'92, ACM, New York, NY, USA, 1992, pp. 83--94.
\newblock \href {http://dx.doi.org/10.1145/143165.143184}
  {\path{doi:10.1145/143165.143184}}.

\bibitem{Pierce2002TAPL}
B.~C. Pierce, Types and Programming Languages, MIT Press, Cambridge, MA, USA,
  2002.

\bibitem{Sangiorgi2012anintroduction}
D.~Sangiorgi, Introduction to Bisimulation and Coinduction, Cambridge
  University Press, New York, NY, USA, 2011.

\bibitem{Coquand1988thecalculus}
T.~Coquand, G.~Huet, The calculus of constructions, Inf. Comput. 76~(2–3)
  (1988) 95--120.
\newblock \href {http://dx.doi.org/10.1016/0890-5401(88)90005-3}
  {\path{doi:10.1016/0890-5401(88)90005-3}}.

\bibitem{Pierce2004ATTAPL}
B.~C. Pierce, Advanced Topics in Types and Programming Languages, The MIT
  Press, Cambridge, MA, USA, 2004.

\bibitem{Seldin1997ontheproof}
J.~P. Seldin, On the proof theory of {Coquand}'s calculus of constructions,
  Annals of Pure and Applied Logic 83~(1) (1997) 23--101.
\newblock \href {http://dx.doi.org/10.1016/S0168-0072(96)00008-5}
  {\path{doi:10.1016/S0168-0072(96)00008-5}}.

\bibitem{Pierce2013SF}
B.~C. Pierce, C.~Casinghino, M.~Greenberg, C.~Hri\c{t}cu, V.~Sj\"{o}berg,
  B.~Yorgey, \href{http://cis.upenn.edu/~bcpierce/sf}{Software Foundations},
  2013, electronic textbook.
\newline\urlprefix\url{http://cis.upenn.edu/~bcpierce/sf}

\bibitem{Ciobaca2013fromsmallstep}
{\cb S}.~Ciob\^ac\u{a}, From small-step semantics to big-step semantics,
  automatically, in: IFM'13, Vol. 7940 of LNCS, Springer, 2013, pp. 347--361.
\newblock \href {http://dx.doi.org/10.1007/978-3-642-38613-8_24}
  {\path{doi:10.1007/978-3-642-38613-8_24}}.

\bibitem{Churchill2015reusable}
M.~Churchill, P.~D. Mosses, N.~Sculthorpe, P.~Torrini, Reusable components of
  semantic specifications, in: Trans. Aspect-Oriented Software Development XII,
  Vol. 8989 of LNCS, Springer, 2015, pp. 132--179.
\newblock \href {http://dx.doi.org/10.1007/978-3-662-46734-3_4}
  {\path{doi:10.1007/978-3-662-46734-3_4}}.

\bibitem{Churchill2013modularbisimulationtheory}
M.~Churchill, P.~D. Mosses, Modular bisimulation theory for computations and
  values, in: FoSSaCS'13, Vol. 7794 of LNCS, Springer, 2013, pp. 97--112.
\newblock \href {http://dx.doi.org/10.1007/978-3-642-37075-5_7}
  {\path{doi:10.1007/978-3-642-37075-5_7}}.

\bibitem{Milner1997thedefinition}
R.~Milner, M.~Tofte, R.~Harper, D.~MacQueen, The Definition of Standard ML, MIT
  Press, Cambridge, MA, USA, 1997.

\bibitem{BachPoulsen2014derivingpretty}
C.~Bach~Poulsen, P.~D. Mosses, Deriving pretty-big-step semantics from
  small-step semantics, in: ESOP 2014, Vol. 8410 of LNCS, Springer, 2014, pp.
  270--289.
\newblock \href {http://dx.doi.org/10.1007/978-3-642-54833-8_15}
  {\path{doi:10.1007/978-3-642-54833-8_15}}.

\bibitem{Nakata2009tracebased}
K.~Nakata, T.~Uustalu, Trace-based coinductive operational semantics for
  {While}, in: TPHOLs'09, Vol. 5674 of LNCS, Springer, 2009, pp. 375--390.
\newblock \href {http://dx.doi.org/10.1007/978-3-642-03359-9_26}
  {\path{doi:10.1007/978-3-642-03359-9_26}}.

\bibitem{Danielsson2012operationalsemantics}
N.~A. Danielsson, Operational semantics using the partiality monad, in:
  ICFP'12, ACM, New York, NY, USA, 2012, pp. 127--138.
\newblock \href {http://dx.doi.org/10.1145/2364527.2364546}
  {\path{doi:10.1145/2364527.2364546}}.

\bibitem{Abel2014normalization}
A.~Abel, J.~Chapman, Normalization by evaluation in the delay monad: A case
  study for coinduction via copatterns and sized types, in: MSFP'14, Vol. 153
  of ENTCS, Open Publishing Association, 2014, pp. 51--67.
\newblock \href {http://dx.doi.org/10.4204/EPTCS.153.4}
  {\path{doi:10.4204/EPTCS.153.4}}.

\bibitem{Pirog2014thecoinductive}
M.~Pir\'{o}g, J.~Gibbons, The coinductive resumption monad, ENTCS 308 (2014)
  273--288.
\newblock \href {http://dx.doi.org/10.1016/j.entcs.2014.10.015}
  {\path{doi:10.1016/j.entcs.2014.10.015}}.

\bibitem{Hancock2000interactiveprograms}
P.~Hancock, A.~Setzer, Interactive programs in dependent type theory, in:
  CSL'00, Vol. 1862 of LNCS, Springer, 2000, pp. 317--331.
\newblock \href {http://dx.doi.org/10.1007/3-540-44622-2_21}
  {\path{doi:10.1007/3-540-44622-2_21}}.

\bibitem{Bove2009abrief}
A.~Bove, P.~Dybjer, U.~Norell, A brief overview of {Agda} -- a functional
  language with dependent types, in: TPHOLs'09, Springer, 2009, pp. 73--78.
\newblock \href {http://dx.doi.org/10.1007/978-3-642-03359-9_6}
  {\path{doi:10.1007/978-3-642-03359-9_6}}.

\bibitem{Nakata2010resumptions}
K.~Nakata, T.~Uustalu, Resumptions, weak bisimilarity and big-step semantics
  for {While} with interactive {I/O}: An exercise in mixed
  induction-coinduction, in: SOS'10, Vol.~32 of EPTCS, 2010, pp. 57--75.
\newblock \href {http://dx.doi.org/10.4204/EPTCS.32.5}
  {\path{doi:10.4204/EPTCS.32.5}}.

\bibitem{Hur2013thepower}
C.~Hur, G.~Neis, D.~Dreyer, V.~Vafeiadis, The power of parameterization in
  coinductive proof, in: {POPL}'13, {ACM}, New York, NY, USA, 2013, pp.
  193--206.
\newblock \href {http://dx.doi.org/10.1145/2429069.2429093}
  {\path{doi:10.1145/2429069.2429093}}.

\bibitem{Abel2010miniagda}
A.~Abel, {MiniAgda}: Integrating sized and dependent types, in: {PAR}'10,
  Vol.~43 of {EPTCS}, 2010, pp. 14--28.
\newblock \href {http://dx.doi.org/10.4204/EPTCS.43.2}
  {\path{doi:10.4204/EPTCS.43.2}}.

\bibitem{Milner1973processes}
R.~Milner, Processes: A mathematical model of computing agents, in: Logic
  Colloquium '73, Studies in Logic and the Foundations of Mathematics,
  North-Holland Pub. Co., 1975, pp. 157--153.

\bibitem{Uustalu2013coinductivebigstep}
T.~Uustalu, Coinductive big-step semantics for concurrency, in: PLACES'13, Vol.
  137 of EPTCS, 2013, pp. 63--78.
\newblock \href {http://dx.doi.org/10.4204/EPTCS.137.6}
  {\path{doi:10.4204/EPTCS.137.6}}.

\bibitem{Owens2016functionalbigstep}
S.~Owens, M.~O. Myreen, R.~Kumar, Y.~K. Tan, Functional big-step semantics, in:
  ESOP'16, Vol. 9632 of LNCS, Springer, 2016, to appear.

\bibitem{Moggi1991notionsofcomputation}
E.~Moggi, Notions of computation and monads, Inf. Comput. 93~(1) (1991) 55--92.
\newblock \href {http://dx.doi.org/10.1016/0890-5401(91)90052-4}
  {\path{doi:10.1016/0890-5401(91)90052-4}}.

\bibitem{Cenciarelli1993asyntactic}
P.~Cenciarelli, E.~Moggi, A syntactic approach to modularity in denotational
  semantics, in: Category Theory and Computer Science, Proc. 5th Intl. Conf.,
  1993.

\bibitem{BachPoulsen2015imperativepolymorphism}
C.~Bach~Poulsen, P.~D. Mosses, P.~Torrini, Imperative polymorphism by
  store-based types as abstract interpretations, in: PEPM'15, ACM, New York,
  NY, USA, 2015, pp. 3--8.
\newblock \href {http://dx.doi.org/10.1145/2678015.2682545}
  {\path{doi:10.1145/2678015.2682545}}.

\bibitem{BachPoulsen2016xtss}
C.~Bach~Poulsen, Extensible transition system semantics, Ph.D. thesis, Swansea
  University, to appear (2016).

\bibitem{Bahr2015calculating}
P.~Bahr, G.~Hutton, Calculating correct compilers, J. Functional Program. 25
  (2015) e14 (47 pages).
\newblock \href {http://dx.doi.org/10.1017/S0956796815000180}
  {\path{doi:10.1017/S0956796815000180}}.

\bibitem{Leroy2006formalcertification}
X.~Leroy, Formal certification of a compiler back-end or: Programming a
  compiler with a proof assistant, in: POPL'06, ACM, New York, NY, USA, 2006,
  pp. 42--54.
\newblock \href {http://dx.doi.org/10.1145/1111037.1111042}
  {\path{doi:10.1145/1111037.1111042}}.

\bibitem{Wright1994asyntactic}
A.~K. Wright, M.~Felleisen, A syntactic approach to type soundness, Inf.
  Comput. 115~(1) (1994) 38--94.
\newblock \href {http://dx.doi.org/10.1006/inco.1994.1093}
  {\path{doi:10.1006/inco.1994.1093}}.

\bibitem{Harper2000atypetheoretic}
R.~Harper, C.~Stone, A type-theoretic interpretation of {S}tandard {ML}, in:
  G.~Plotkin, C.~Stirling, M.~Tofte (Eds.), Proof, Language, and Interaction,
  MIT Press, Cambridge, MA, USA, 2000, pp. 341--387.

\bibitem{Cousot1997typesas}
P.~Cousot, Types as abstract interpretations, in: POPL'97, ACM, New York, NY,
  USA, 1997, pp. 316--331.
\newblock \href {http://dx.doi.org/10.1145/263699.263744}
  {\path{doi:10.1145/263699.263744}}.

\bibitem{Ager2003afunctional}
M.~S. Ager, D.~Biernacki, O.~Danvy, J.~Midtgaard, A functional correspondence
  between evaluators and abstract machines, in: PPDP'03, ACM, New York, NY,
  USA, 2003, pp. 8--19.
\newblock \href {http://dx.doi.org/10.1145/888251.888254}
  {\path{doi:10.1145/888251.888254}}.

\bibitem{Danvy2008ontheequivalence}
O.~Danvy, K.~Millikin, On the equivalence between small-step and big-step
  abstract machines: A simple application of lightweight fusion, Inf. Process.
  Lett. 106~(3) (2008) 100--109.
\newblock \href {http://dx.doi.org/10.1016/j.ipl.2007.10.010}
  {\path{doi:10.1016/j.ipl.2007.10.010}}.

\bibitem{Simmons2013alogical}
R.~J. Simmons, I.~Zerny, A logical correspondence between natural semantics and
  abstract machines, in: PPDP'13, ACM, New York, NY, USA, 2013, pp. 109--119.
\newblock \href {http://dx.doi.org/10.1145/2505879.2505899}
  {\path{doi:10.1145/2505879.2505899}}.

\end{thebibliography}
\bibliographystyle{elsarticle-num}

\end{document}